\documentclass[fleqn,usenatbib,usedcolumn]{mnras}

\usepackage[T1]{fontenc}
\usepackage{ae,aecompl}

\usepackage[british]{babel}             
\usepackage{graphicx}	
\usepackage{amsmath}	
\usepackage{amssymb}	

\usepackage{color}			   
\definecolor{midgray}{gray}{0.4}		
\definecolor{orange}{rgb}{1,0.5,0}  
\usepackage{breakcites}

\newcommand{\simgt}{\,\rlap{\lower 3.5 pt \hbox{$\mathchar \sim$}} \raise
1pt \hbox {$>$}\,}
\newcommand{\simlt}{\,\rlap{\lower 3.5 pt \hbox{$\mathchar \sim$}} \raise
1pt \hbox {$<$}\,}

\newcommand{\Ob}{\Omega_{\textrm b}}
\newcommand{\Om}{\Omega_{\textrm m}}
\newcommand{\OL}{\Omega_{\Lambda}}

\newcommand{\lya}{Ly$\alpha$}

\newcommand{\xHI}{{\overline{x}_\textrm{{\textsc{hi}}}}}
\newcommand{\DV}{{\Delta v}}

\newcommand{\MUV}{M_\textsc{uv}}

\newcommand{\OII}{[O\textsc{ii}]}
\newcommand{\OIII}{O\textsc{iii}}
\newcommand{\CIV}{C\textsc{iv}}
\newcommand{\HeII}{He\textsc{ii}}

\newcommand{\fdens}{erg s$^{-1}$ cm$^{-2}$}
\newcommand{\kms}{km s$^{-1}$}

\newcommand{\HST}{\textit{HST}}

\newcommand{\BE}{\begin{equation}}
\newcommand{\EE}{\end{equation}}
\newcommand{\BEA}{\begin{align}}
\newcommand{\EEA}{\end{align}}

\defcitealias{Mason2018a}{M18a}	

\title[KLASS $z\sim8$]{Inferences on the Timeline of Reionization at $z\sim8$\\From the KMOS Lens-Amplified Spectroscopic Survey}

\author[C. A. Mason et al.]{Charlotte~A.~Mason$^{1,2}$\thanks{Hubble Fellow}\thanks{E-mail: charlotte.mason@cfa.harvard.edu},
Adriano~Fontana$^{3}$,
Tommaso~Treu$^{1}$,
Kasper~B.~Schmidt$^{4}$,\newauthor
Austin~Hoag$^{1,5}$,
Louis~Abramson$^{6}$,
Ricardo~Amorin$^{7,8}$,
Maru\v{s}a~Brada\v{c}$^{5}$,\newauthor
Lucia~Guaita$^{3,9}$,
Tucker~Jones$^{5}$,
Alaina~Henry$^{10}$,
Matthew~A.~Malkan$^{1}$,\newauthor
Laura~Pentericci$^{3}$,
Michele~Trenti$^{11,12}$,
and
Eros~Vanzella$^{13}$
\\
$^{1}$ Department of Physics and Astronomy, UCLA, Los Angeles, CA, 90095-1547, USA\\
$^{2}$ Center for Astrophysics | Harvard \& Smithsonian, 60 Garden St, Cambridge, MA, 02138, USA\\
$^{3}$ INAF Osservatorio Astronomico di Roma, Via Frascati 33, I-00040 Monteporzio (RM), Italy\\
$^{4}$ Leibniz-Institut f\"{u}r Astrophysik Potsdam (AIP), An der Sternwarte 16, 14482 Potsdam, Germany\\
$^{5}$ Department of Physics, University of California, Davis, CA, 95616, USA\\
$^{6}$ Carnegie Observatories, 813 Santa Barbara St., Pasadena, California 91101-1292, USA\\
$^{7}$ Instituto de Investigaci\'on Multidisciplinar en Ciencia y Tecnolog\'ia, Universidad de La Serena, Ra\'ul Bitr\'an 1305, La Serena, Chile\\
$^{8}$ Departamento de F\'isica y Astronom\'ia, Universidad de La Serena, Av. Juan Cisternas 1200 Norte, La Serena, Chile\\
$^{9}$ N\'{u}cleo de Astronom\'{i}a, Facultad de Ingenier\'{i}a, Universidad Diego Portales, Av. Ej\'{e}rcito 441, Santiago, Chile \\
$^{10}$ Space Telescope Science Institute, 3700 San Martin Drive, Baltimore, MD 21218\\
$^{11}$ School of Physics, University of Melbourne, Parkville, Victoria, Australia\\
$^{12}$ ARC Centre of Excellence for All Sky Astrophysics in 3 Dimensions (ASTRO 3D)\\
$^{13}$ INAF -- OAS, Osservatorio di Astrofisica e Scienza dello Spazio di Bologna, via Gobetti 93/3, I-40129 Bologna, Italy
}

\date{Accepted XXX. Received YYY; in original form ZZZ}

\pubyear{2019}

\begin{document}
\label{firstpage}
\pagerange{\pageref{firstpage}--\pageref{lastpage}}
\maketitle

\begin{abstract}
Detections and non-detections of Lyman alpha (\lya) emission from $z>6$ galaxies ($<1$ Gyr after the Big Bang) can be used to measure the timeline of cosmic reionization. Of key interest to measuring reionization's mid-stages, but also increasing observational challenge, are observations at $z>7$, where \lya\ redshifts to near infra-red wavelengths. Here we present a search for $z>7.2$ \lya\ emission in 53 intrinsically faint Lyman Break Galaxy candidates, gravitationally lensed by massive galaxy clusters, in the KMOS Lens-Amplified Spectroscopic Survey (KLASS). With integration times of $\sim7-10$ hours, we detect no \lya\ emission with S/N $>5$ in our sample. We determine our observations to be 80\% complete for $5\sigma$ spatially and spectrally unresolved emission lines with integrated line flux $>5.7\times10^{-18}$\,\fdens. We define a photometrically selected sub-sample of 29 targets at $z=7.9\pm0.6$, with a median $5\sigma$ \lya\ EW limit of 58\,\AA. We perform a Bayesian inference of the average intergalactic medium (IGM) neutral hydrogen fraction using their spectra. Our inference accounts for the wavelength sensitivity and incomplete redshift coverage of our observations, and the photometric redshift probability distribution of each target. These observations, combined with samples from the literature, enable us to place a lower limit on the average IGM neutral hydrogen fraction of $> 0.76 \; (68\%), \; > 0.46 \; (95\%)$ at $z\sim8$, providing further evidence of rapid reionization at $z\sim6-8$. We show that this is consistent with reionization history models  extending the galaxy luminosity function to $\MUV \simlt -12$, with low ionizing photon escape fractions, $f_\textrm{esc} \simlt 15\%$.
\end{abstract}

\begin{keywords}
dark ages, reionization, first stars -- galaxies: high-redshift -- galaxies: evolution -- intergalactic medium
\end{keywords}


\section{Introduction}
\label{sec:intro}


The reionization of intergalactic hydrogen in the universe's first billion years is likely linked to the formation of the first stars and galaxies: considered to be the primary producers of hydrogen-ionizing photons \citep[e.g.,][]{Lehnert2003,Bouwens2003,Yan2004,Bunker2004,Shull2012,Bouwens2015a}. Accurately measuring the timeline of reionization enables us to constrain properties of these first sources \citep[e.g.,][]{Robertson2013,Robertson2015,Greig2017}.

Measurements of the reionization timeline are challenging, however, due to the rarity of bright quasars at $z>6$ \citep{Fan2001,Manti2016,Parsa2018}, which have historically provided strong constraints on the end stages of reionization \citep[e.g.,][]{Fan2006,McGreer2014,Greig2016,Banados2017}. In the coming decade 21\,cm observations are expected to provide information about the $z>6$ IGM and the nature of the first galaxies \citep[e.g.,][]{Liu2016,Mirocha2016}, but current progress has been driven by observations of \lya\ (rest-frame 1216\,\AA) emission in galaxies, using near infra-red (NIR) spectroscopy.

\lya\ is a highly resonant line, and strongly scattered by intervening neutral hydrogen as it travels to our telescopes. Whilst young star-forming galaxies, selected with a Lyman Break (Lyman Break Galaxies -- LBGs) show \lya\ emission in abundance up to $z\sim6$ \citep[e.g.,][]{Stark2011,Hayes2011,Curtis-Lake2012,Cassata2015,DeBarros2017}, at higher redshifts the fraction of galaxies detected with \lya\ emission, and the scale length of the \lya\ rest-frame equivalent width (EW) distribution, decreases rapidly \citep[e.g.,][]{Fontana2010,Pentericci2011,Caruana2012,Treu2012,Treu2013,Ono2012,Caruana2014,Pentericci2014,Schenker2014,Tilvi2014,Faisst2014,Jung2018b}. This rapid decline of detected \lya\ emission is most plausibly due to absorption in an increasingly neutral IGM \citep{Dijkstra2011,Dijkstra2014a,Mesinger2015}. 

Large spectroscopic surveys of LBG candidates are being assembled out to $z\sim7$ \citep{Pentericci2011,Pentericci2014,Pentericci2018,Hoag2019} but exploring the earliest stages of reionization requires us to observe \lya\ at even higher redshifts. Only a handful of \lya\ emitters have been confirmed at $z\simgt7.5$ \citep{Zitrin2015a,Oesch2015,Roberts-Borsani2016,Stark2017,Hoag2017}, where the dominance of sky emission in the NIR makes observations of faint sources even more challenging. Additionally, because \lya\ emission can be spatially extended and/or offset from the UV continuum emission \citep{Wisotzki2016,Leclercq2017}, it is likely that slit-based spectroscopy is not capturing the full \lya\ flux. Hence, the observed decline in \lya\ emission at $z>6$ could be partially due to redshift-dependent slit-losses as well as reionization.

In this paper we present a search for $z\simgt7.2$ \lya\ emission in NIR spectroscopy of 53 intrinsically faint LBG candidates ($\MUV \simgt -20$), gravitationally lensed behind 6 massive galaxy clusters, including 4 of the Frontier Fields \citep{Lotz2017}, selected from the Grism Lens-Amplified Survey from Space \citep[hereafter GLASS,][]{Schmidt2014,Treu2015}. We also present observations of \CIV\ emission in 3 images of a previously confirmed multiply-imaged $z=6.11$ galaxy \citep{Boone2013,Balestra2013,Monna2014}. 

The observations presented in this work were carried out with the ESO Very Large Telescope (\textit{VLT}) K-band Multi Object Spectrometer \citep[hereafter KMOS,][]{Sharples2013}. This work presents the first results of $z>3.8$ observations with KMOS. KMOS is an integral field unit (IFU) instrument, and we demonstrate here that our observations are more complete to spatially extended and/or offset \lya\ emission than traditional slit spectrographs. 

We use our new deep spectroscopic observations to infer the average IGM neutral hydrogen fraction ($\xHI$) at $z\sim8.$ \citet[][hereafter M18a]{Mason2018a} presented a flexible Bayesian framework to directly infer $\xHI$ from detections and non-detections of \lya\ emission from LBGs. The framework combines realistic inhomogeneous reionization simulations and models of galaxy properties. That work measured $\xHI = 0.59_{-0.15}^{+0.11}$ ($16-84\%$ confidence intervals) at $z\sim7$. Building on \citet{Treu2012} and \citetalias{Mason2018a} we extend this framework to use the full spectra obtained in our observations for the Bayesian inference, accounting for the incomplete wavelength coverage and spectral variation of the noise, and marginalising over emission linewidth. Our framework uses the photometric redshift probability distribution, obtained from deep photometry including new Spitzer/IRAC data, of each object to robustly account for uncertainties in redshift determination.

The paper is structured as follows: Section~\ref{sec:obs} describes our KMOS observations and the target selection from the GLASS parent sample; Section~\ref{sec:results} describes the search for \lya\ emission in our KMOS data cubes, and the purity and completeness of our survey; and Section~\ref{sec:reionization} describes the Bayesian inference of the neutral fraction and presents our limit on $\xHI$ at $z\sim8$. We discuss our findings in Section~\ref{sec:dis}, including an assessment of the performance of KMOS for background-limited observations using our deep observations, and summarise in Section~\ref{sec:conc}.

We use the \citet{PlanckCollaboration2015} cosmology where $(\OL, \Om, \Ob, n,  \sigma_8, H_0) =$ (0.69, 0.31, 0.048, 0.97, 0.81, 68 \kms\ Mpc$^{-1}$). All magnitudes are given in the AB system.

\section{Observations}
\label{sec:obs}


\subsection{The KMOS Lens-Amplified Spectroscopic Survey}
\label{sec:obs_KLASS}

KLASS is an ESO VLT KMOS Large Program (196.A-0778, PI: A. Fontana) which targeted the fields of six massive galaxy clusters: Abell 2744 (hereafter A2744); MACS J0416.1-2403 (M0416); MACS J1149.6+2223 (M1149); MACS J2129.4-0741 (M2129); RXC J1347.5-1145 (RXJ1347); and RXC J2248.7-4431 (RXJ2248, aka Abell S1063). A2744, M0416, M1149 and RXJ2248 are all Frontier Fields \citep[hereafter HFF,][]{Lotz2017}. Observations were carried out in Service Mode during Periods $96-99$ (October 2015 - October 2017).

KMOS is a multi-object IFU spectrograph, with 24 movable IFUs, split between 3 different spectrographs \citep{Sharples2013}. Each IFU is $2\farcs8 \times 2\farcs8$ field of view, with pixel size $0\farcs2 \times 0\farcs2$, and 2048 pixels along the wavelength axis\footnote{We use the following definitions for describing 3D spectra in this paper. Pixel: 2D spatial pixel (size $0\farcs2 \times 0\farcs2$). Spaxel: the 1D spectrum in a single spatial pixel (spanning the spectral range $\sim1-1.35\,\mu$m, in 2048 spectral pixels). Voxel: 3D pixel in the data cube with both spatial and spectral indices.}.

The key science drivers of KLASS are:

\begin{enumerate}
\item To probe the internal kinematics of galaxies at $z\sim1-3$, with superior spatial resolution compared to surveys in blank fields \citep{Mason2017}.
\item To investigate $z\simgt7$ \lya\ emission from the GLASS sample, independently of the HST spectroscopic observations, providing validation and cross-calibration of the results and enabling us to constrain the timeline and topology of reionization \citep{Treu2012,Treu2013,Schmidt2016,Mason2018a}.
\end{enumerate}


\citet{Mason2017} addressed the first science driver by presenting spatially resolved kinematics in 4 of the 6 KLASS clusters from our early data, including five of the lowest mass galaxies with IFU kinematics at $z>1$, and provided evidence of mass-dependent disk settling at high redshift \citep{Simons2017}. The KLASS kinematic data were combined with metallicity gradients from the \HST\ GLASS data to enable the study of metallicity gradients as a diagnostic of gas inflows and outflows \citep{Wang2016}.

This paper addresses the second science driver by presenting our $z>7$ candidate targets with complete exposures. We use the YJ observing band, giving us access to \lya\ emission at $z\sim7.2-10.1$.

The choice of an IFU instrument for high-redshift \lya\ observations was motivated by indications that ground-based slit-spectroscopy measures lower \lya\ flux than \HST\ slit-less grism spectroscopy \citep{Tilvi2016,Huang2016a,Hoag2017}, which, as well as reionization, could contribute to the observed decline in \lya\ emission at $z>6$. \lya\ emission can be spatially extended and/or offset from the UV continuum emission \citep{Feldmeier2013,Momose2014,Wisotzki2016,Leclercq2017}, so it is likely that slit-based spectrographs do not capture the full \lya\ flux.

By using IFUs our observations should be more complete to spatially extended and/or offset \lya\ than traditional slit spectrographs. \citet{Mason2017} showed that only $\sim60\%$ of emission line flux was contained in $\sim0\farcs7$ simulated slits \citep[a typical slit-width used for \lya\ observations, e.g.,][]{Hoag2017} on KMOS spectra, whereas the full flux is captured within the $2\farcs8 \times 2\farcs8$ KMOS field of view. Thus we expect most \lya\ flux to be captured within the KMOS IFUs. The $2\farcs8$ wide IFUs cover $\sim14$ proper kpc at $z\sim8$, while the UV effective radii of galaxies at these redshifts is only $\simlt 1$ proper kpc \citep{Shibuya2015}. We demonstrate in Section~\ref{sec:results_completeness} that our KMOS observations have good completeness for spatially extended and/or offset \lya\ emission.

\subsection{Target selection}
\label{sec:obs_targets}

\begin{table*}
\centering
\caption{KLASS cluster targets}
\label{tab:targets}
\begin{tabular}[c]{lcccccc}
\hline
\hline
Cluster & Run ID & DIT [s]& NDITs$^\ast$ & Exposure [hrs]	& \multicolumn{2}{c}{Number of targets}\\
		& 		 &        & 		  &  				&	Category 1	& Category 2	\\
\hline	
A2744$^\dagger,^\ddagger$	& A 	 & 900 & 25 & 6.25  		&	3	& 7	\\
M0416$^\ddagger$		& B 	 & 900 & 43 & 10.75 		&	2	& 5 \\
M1149$^\ddagger$		& C 	 & 900 & 40 & 10.00 		&	2	& 6 \\
M2129 		& E 	 & 450 & 85 & 10.625		&	3	& 7	\\
RXJ1347 	& D 	 & 450 & 88 & 11.00 		&	3	& 8	\\
RXJ2248$^\diamond$ & F 	 & 300 & 93 & 7.75 			&	1	& 6	\\
\hline
\multicolumn{7}{p{.7\textwidth}}{\textsc{Note.} -- $^\ast$ The number of Detector Integration Times (DITs) used in this analysis: we discarded DITs if the seeing was $>0\farcs8$ as measured by stars observed in each DIT. The total exposure time = DIT $\times$ NDITs. $^\dagger$ We had to discard our initial 4 hours of observations of A2744 due to irreparable flexure issues due to rotating the instrument between science and sky DITs. All subsequent observations were performed with no rotation of the instrument between science and sky DITs. $^\ddagger$ Target selection in these clusters was primarily done from preliminary versions of the ASTRODEEP catalogues \citep{Castellano2016b,Merlin2016,DiCriscienzo2017}, which did not include Spitzer/IRAC photometry. $^\diamond$ Due to a high proper motion reference star, some of the observations of RXJ2248 were taken at a slight offset from the required target centre, reducing the total exposure at that position. RXJ2248 also included 3 $z=6.11$ targets (Appendix~\ref{app:CIV}).}
\end{tabular}
\end{table*}
KLASS targets were selected from the GLASS survey\footnote{\url{http://glass.astro.ucla.edu}} \citep{Schmidt2014,Treu2015}, a large Hubble Space Telescope (\HST) slit-less grism spectroscopy program. GLASS obtained spectroscopy of the fields of 10 massive galaxy clusters, including the HFF and 8 CLASH clusters \citep{Postman2012}. The Wide Field Camera 3 (WFC3) grisms G102 and G141 were used to cover the wavelength range $0.8 - 1.6\,\mu$m with spectral resolution $R\sim150$. We refer the reader to \citet{Schmidt2014} and \citet{Treu2015} for full details of GLASS.

KLASS observations aimed to provide the high spectral resolution necessary to measure the purity and completeness of the grism spectra, to measure lines that were unresolved in \HST, and to obtain velocity information for $z\sim1$ targets which the low resolution grisms cannot provide. In combination with additional GLASS follow-up observations at Keck \citep{Huang2016a,Hoag2017,Hoag2019} we will address the purity and completeness of the HST grisms in a future work. In this work we present our high-redshift candidate targets and our inferences about reionization obtained from the KLASS data.

Two categories of high-redshift candidate KLASS targets were selected from the GLASS data:

\begin{enumerate}
\item Category 1: 14 objects with marginal (S/N $\sim3$) candidate \lya\ emission in the \HST\ grisms, identified by visual inspection of the GLASS data, which fall within the KMOS YJ spectral coverage ($\sim1-1.35\,\mu$m). 4 candidates were selected from a list of candidates in a preliminary census of GLASS data by \citet{Schmidt2016}. The remaining candidates were selected in a similar method to the procedure followed by \citet{Schmidt2016}.
\item Category 2: 39 LBG candidates selected with $z_\textrm{phot} > 7.2$, from an ensemble of photometric catalogues described by \citet{Schmidt2016}. This includes three LBGs which were spectroscopically confirmed via sub-mm emission lines after our survey began: A2744\_YD4 (A2744\_2248 in this paper), at $z=8.38$ \citep[][discussed in Section~\ref{sec:reionization_phot} and \ref{sec:reionization_infer}]{Laporte2017}, M0416\_Y1 (M0416\_99 in this paper), at $z=8.12$ \citep[][discussed in Section~\ref{sec:reionization_phot}]{Tamura2018}, and M1149\_JD1, at $z=9.11$ \citep[][discussed in Section~\ref{sec:dis_JD1}]{Hashimoto2018}.
\end{enumerate}

An additional three targets were multiple images of the $z=6.11$ system in RXJ2248 \citep{Boone2013,Balestra2013,Monna2014} where we targeted \CIV$\lambda$1548,1551 emission. This object is discussed in Appendix~\ref{app:CIV}.

We ranked objects in order of the number of inspectors who reported a candidate emission line for our Category 1 targets, and then by the number of independent photometric catalogues the target appeared in (for both categories). Our observations were planned prior to the release of the full HFF datasets, so the photometric catalogues we used to select candidates did not contain the full photometry now available. In particular, deep Spitzer/IRAC data did not exist, which can be useful for distinguishing between high-redshift star forming galaxies and $z\sim1-2$ passive galaxies. Nor were sophisticated intra-cluster light (ICL) removal techniques developed at that point \citep[e.g.,][]{Merlin2016,Morishita2017a,Livermore2017}. 

Thus our LBG selection was heterogeneous, but in this paper we now add in the new deep and extended photometry to define a homogeneous photometric selection. We expect some faint candidates may have been spurious in the initial photometry and may not appear in the final deep catalogues. Additionally we expect that with the inclusion of Spitzer/IRAC photometry some of the objects originally selected to be $z>7$ may be low redshift contaminants. In our reionization analysis we use catalogues built using the final HFF datasets to define a selection function for a photometrically-selected sample for our inference (described in Section~\ref{sec:reionization_phot}). We demonstrate that this KLASS sub-sample is not a biased sample of the final parent catalogues in Appendix~\ref{app:phot}.

The GLASS median $1\sigma$ flux limit is $5\times10^{-18}$\,\fdens\ \citep{Schmidt2016}, and we tried to be as inclusive as possible when assigning candidates to the KMOS IFUs from the GLASS parent catalogue. Most of the candidates were only $3\sigma$ significance in GLASS data and we were aiming to provide confirmation of those tentative targets our deep KMOS observations - though for our ground-based observations, at least 50\% of the wavelength range is dominated by sky emission and the low spectral resolution ($R\sim100$) of the HST grisms means that the line position is uncertain by $\pm25$\,\AA\ so the lines could be in bad sky regions. Additionally, in planning our observations we likely overestimated the sensitivity of KMOS YJ using the online exposure time calculator, especially at the blue end of the detectors. We discuss this in more detail in Section~\ref{sec:dis_KMOS}.

As we describe below in Section~\ref{sec:results_flim}, our KLASS observations are $\sim80\%$ complete for lines with flux $>5.7\times10^{-18}$\,\fdens, which suggests we should have confirmed the majority of the 14 GLASS candidate \lya\ emission targets we observed (Category 1, described in Section~\ref{sec:obs_targets}). However, we did not detect any emission in the cubes containing these candidates, suggesting at least some of the GLASS candidates were spurious noise peaks in the HST grisms. A more thorough comparison of the GLASS HST grism and ground-based follow-up observations \citep[including KLASS and Keck observations,][]{Huang2016a,Hoag2017,Hoag2019} to recover the grism purity and completeness will be left to a future work.

53 $z_\mathrm{phot} > 7.2$ candidate targets across the 6 clusters were assigned to 51 KMOS IFUs (two IFUs contained two nearby candidates). The cluster list and number of high-redshift candidate targets per cluster is shown in Table~\ref{tab:targets}.

\subsection{KLASS observing strategy and reduction}
\label{sec:obs_obsplan}

KLASS observations were carried out with KMOS YJ ($\sim1-1.35\,\mu$m). The spectral resolution $R\sim3400$ is sufficient to distinguish \lya\ from potential low redshift contaminants with the \OII$\lambda3726,3729$ emission doublet at $z\sim2$. 

Observations were carried out in service mode and executed in one hour observing blocks with repeating ABA science-sky-science integration units (detector integration times -- DITs). Each observing block comprised 1800\,s of science integration, and 900\,s on sky. Pixel dither shifts were included between science frames. A star was observed in 1 IFU in every observing block to monitor the point spread function (PSF) and the accuracy of dither offsets. The PSF was well-described by a circular Gaussian and the median seeing of our observations was FWHM $\sim0\farcs6$.

In each cluster, the 3 top priority targets were observed for $1.5\times$ the average exposure time by assigning 2 IFUs per target and nodding between them during A and B modes.

\subsection{Reduction}
\label{sec:obs_reduction}

Data were reduced using the ESO KMOS pipeline v.1.4.3 \citep{Davies2013}. We apply a correction for known readout channel level offsets before running the pipeline. We run the pipeline including optimised sky subtraction routines \texttt{sky\_tweak} \citep{Davies2007} and \texttt{sky\_scale}. 

To improve the sky subtraction in the pipeline-reduced `A-B' cubes we produced master sky residual spectra by median combining all IFUs on each spectrograph into a 1D master sky residual spectrum for each DIT, excluding cubes containing $z\simlt2$ targets with bright emission lines and/or continua. We then subtract these 1D sky residual spectra from the `A-B' cubes on the same spectrograph for each DIT, rescaling the 1D spectra in each spaxel so that the resulting 1D spectrum in each spaxel is minimised.

Similar techniques to improve sky subtraction are described by \citet{Stott2016}. This method worked best for our survey design. We note that this method performed better than in-IFU sky residual subtraction (i.e. subtracting a median sky residual spectrum produced from `empty' spaxels in each IFU) as it preserved emission line flux in the modestly sized KMOS IFUs.

Cube frames from each DIT are combined via sigma clipping, using spatial shifts determined by the position of the star observed in the same IFU in each DIT, to produce flux and noise cubes. For this work we used only frames with seeing $\leq 0\farcs8$ (as measured by the star observed in our science frames). The median seeing was $\sim 0\farcs6$. DIT length, observing pattern and total integration times used for this paper are listed in Table~\ref{tab:targets}. We note that due to the failure of one of the KMOS arms, no star was observed in the A2744 observations. We used a bright $z\sim1$ target to estimate the dither offsets for this cluster.

For pure Gaussian noise, the pixel distribution of S/N should be normally distributed. We tested this by selecting non-central regions of cubes containing high-redshift candidate targets (i.e. where we expect very little source flux) and found the pixel distribution of S/N to have standard deviation $>1$, suggesting the noise is underestimated by the pipeline. 

We therefore apply a rescaling to the noise of the combined cubes. We create an average 1D noise spectrum in a single spaxel for each cluster by taking the root-mean-square (RMS) at every wavelength of every spaxel from the cubes containing high-redshift candidate targets. Since the cubes are predominantly noise, taking the RMS of the flux at each wavelength across multiple cubes should give the appropriate noise. We find this RMS spectrum is $\sim1.2\times$ higher than the pipeline average 1D noise spectrum (taking the average of the noise cubes across the same set of high-redshift targets). We rescale the pipeline noise in every cube by this ratio of the cluster RMS noise spectrum to the cluster average noise spectrum. 

Finally, we rescale the noise in each cube by a constant value so that the S/N distribution of all pixels has standard deviation 1 (clipping pixels within 99.9\% to remove spurious peaks). We find the S/N distribution is well-described by a Gaussian distribution, with non-Gaussian tails only beyond the $\simgt7\sigma$ confidence regions, due to bad sky subtraction residuals.

\section{Emission Line Search, Purity and Completeness}
\label{sec:results}

In this section we describe our search for \lya\ emission in our KMOS observations. We give our algorithm for line detection in Section~\ref{sec:results_lines}, and calculate the purity and completeness of our observations in Sections~\ref{sec:results_detections} and~\ref{sec:results_completeness}. Given that we detect no convincing \lya\ emission lines in our sample we present our flux and EW upper limits in Section~\ref{sec:results_flim}.

\subsection{Emission line detection technique}
\label{sec:results_lines}

To search for emission lines in the KMOS cubes, to robustly determine the completeness and purity of our survey, and determine the flux limits of our observations, we used the following algorithm to flag potential lines:
\begin{enumerate}
	\item Create a circular aperture with $r = 2 \sigma_\textsc{psf} \sim 0\farcs5 \sim 2.5$ pixels (using our median seeing $FWHM_\mathrm{psf} = 0\farcs6$), which will capture $86\%$ of the total flux for spatially unresolved emission line at the centre of the aperture.
	\item Sum the flux, and take the RMS of the noise of all spaxels in the aperture to create 1D data and noise spectra.
	\item Rescale the 1D noise spectrum so the S/N in all pixels (excluding the 0.1\% most extreme S/N values) is Normal.
	\item Scan through in wavelength and flag a detection if 3 adjacent wavelength pixels have $S/N > 3$. This corresponds to a $S/N \simgt 5$ detection of the integrated line flux.
	\item Iterate over 25 apertures centred within 3 pixels ($0\farcs6$) of the IFU centre, i.e. $x = [-3,-1.5,0,1.5,3]$, $y = [-3,-1.5,0,1.5,3]$ where $(x,y) = (0,0)$ is the IFU centre.
\end{enumerate}
Our search covers $\sim 25 \times 2000 = 50,000$ potential emission line positions in each cube. As our detection threshold is $5\sigma$ we would expect a false positive rate of $6 \times 10^{-7}$, i.e. $\sim0.03$ false detections per cube for Gaussian noise. As discussed in Section~\ref{sec:obs_reduction} the S/N has small non-Gaussian tails due to sky subtraction residuals so we expect a slightly higher false detection rate than this.

\subsection{Candidate emission lines and sample purity}
\label{sec:results_detections}

We ran the detection algorithm described in Section~\ref{sec:results_lines} on the 54 cubes containing our high-redshift candidate targets (including the 3 cubes containing the $z=6.11$ images). 9 unique candidate lines were flagged (combining candidates at the same wavelength identified in different apertures). Each of these candidate lines was then visually inspected to determine whether it was a true emission line or a spurious noise peak. For our inspections we use both 1D spectra extracted in the detection apertures as well as 2D collapsed images of the candidate line obtained by summing cube voxels in the wavelength direction. The 2D images are helpful for determining plausible spatially compact emission from the uniform emission produced by sky residuals.

Our algorithm correctly identifies the \CIV$\lambda1551$ emission at 11023.7\,\AA\ in the brightest image of the multiply-imaged $z=6.11$ system, demonstrating the depth of our KMOS observations and the fidelity of our algorithm. Another detection is flagged in this object at 13358.6\,\AA\ but the emission appears diffuse and the wavelength is not consistent with other expected UV emission lines so we deem this spurious. We describe this object in more detail in Appendix~\ref{app:CIV}. 

Of the remaining 7 lines flagged, 6 are deemed to be spurious detections as they are at the spectral edges of the detector, or immediately adjacent to strong skylines and appear to have P-Cygni profiles, indicating extreme sky subtraction failures. Whilst it could be possible to add a cut to e.g. downweight flagged lines adjacent to skylines, given the relatively low spectral resolution of our observation ($R\sim3400$) we were wary that many true emission lines could be overlapping with skylines, thus visual inspection was necessary. This is clearly demonstrated in our detection of \CIV\ emission where both doublet components overlap with sky lines (see Figure~\ref{fig:CIV}).

The remaining candidate emission line at 12683.7\,\AA\ is spatially offset from the $z>7$ LBG candidate in the cube. We determine the detected emission to be associated with a nearby ($\sim1.1\arcsec$) galaxy with $z_\textrm{phot} = 4.2$, which has bright continuum emission in the GLASS data. The candidate line appears in a particularly bad spectral region of telluric absorption, and we determine the detection to be due to inadequate continuum subtraction of the $z\sim4$ source.

In our reductions we subtract a sky residual spectrum to minimise the flux in each spaxel of the high-redshift candidate cubes (Section~\ref{sec:obs_reduction}). During that process most of the continuum emission from the $z\sim4$ object was poorly subtracted by scaling the sky residual spectrum to high values. Some residual flux is left, which correlates with the positions of sky residuals. We note that the LBG candidate targeted in this IFU is not present in the final deep photometric catalogues and is excluded from our reionization (it was likely a spurious detection in the original shallow photometry, Section~\ref{sec:reionization_phot}). We remove this cube from further analysis. 

Thus we determine our algorithm has detected 1 real emission line, and 7 spurious detections (excluding the $z\sim4$ continuum object described above), allowing us to define the purity of our spectral sample:
\BE \label{eqn:purity}
P = 1 - \frac{N_\textrm{spurious}}{N_\textrm{pos}}
\EE
where $N_\textrm{spurious} = 17$ is the total number of spurious flags (8 unique false detections which were sometimes flagged in multiple apertures) and $N_\textrm{pos} = 101763 \times 25$ is the number of possible emission line positions in the 53 useful cubes, removing wavelength pixels not covered by certain detectors, in 25 apertures. We measure $P=1-7\times10^{-6}$. Our spurious detection rate is $\sim10\times$ higher than that expected for $5\sigma$ fluctuations in the noise, which was expected due to the non-Gaussian tail in our S/N distribution due to sky subtraction residuals. To verify that the S/N distribution is symmetrical we also ran the detection algorithm to look for negative peaks (S/N$\simlt-5$) which should occur at the same rate. We found 12 flagged negative S/N detections, comparable to our 7 flagged spurious detections with positive S/N.

We ran the algorithm on our Category 1 sources with a lower S/N threshold: S/N $>2.5$ per wavelength pixel, corresponding to $S/N \simgt 4$ in the integrated line. We found no convincing detections with this lower threshold and are thus unable to confirm any of the candidate GLASS emission lines. Given that most of the GLASS \lya\ candidates were of low significance in the GLASS \HST\ data these candidates may have been spurious noise peaks in the grism data.

In Section~\ref{sec:reionization_phot} below we list the \lya\ flux and EW limits for our most likely $z_\mathrm{phot}$ LBGs candidates. We discuss our limits on other UV lines in Section~\ref{sec:dis_otherUV}.

\subsection{Completeness}
\label{sec:results_completeness}

\begin{figure*} 
\centering
\includegraphics[width=0.8\textwidth, clip=true]{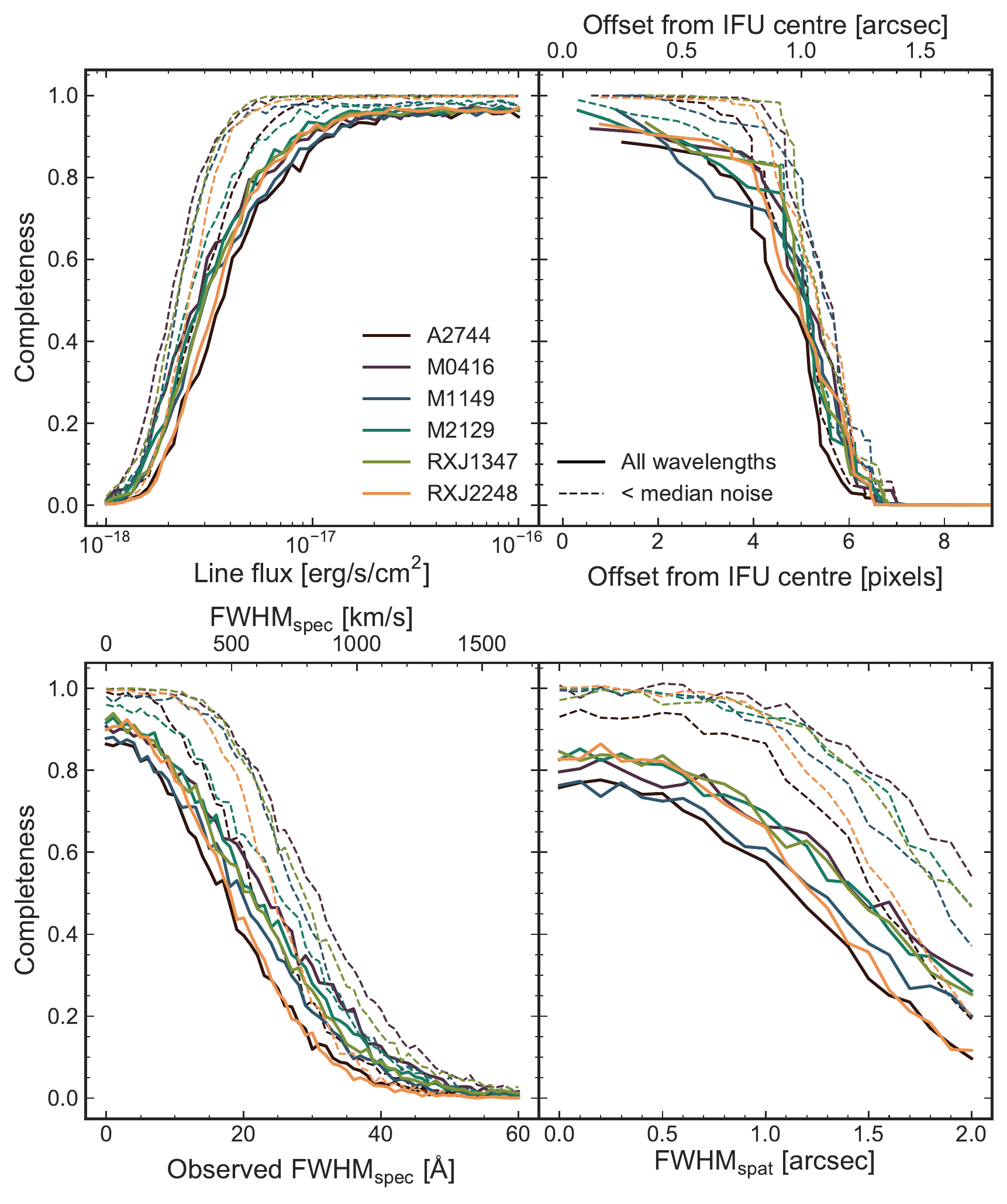}
\caption{Completeness as a function of line flux (\textbf{top left}), spatial offset from IFU centre (\textbf{top right}), spectral linewidth (\textbf{lower left}) and spatial extent (\textbf{lower right}). Each colour corresponds to a separate cluster target. Dashed lines show the completeness across the entire wavelength range, solid lines show the completeness in wavelength regions where the noise level is below the median. FWHM$_\textrm{spec}$ velocities were calculated assuming $z=8$. Spatial extent, FWHM$_\textrm{spat}$ is the extent of the source (excluding the PSF). We create simulated lines with total spatial extent $\textrm{FWHM}_\textrm{spat,tot} = \sqrt{\textrm{FWHM}_\textrm{PSF}^2 + \textrm{FWHM}_\textrm{spat}^2}$. In each plot the parameter of interest and wavelength are varied, while the other parameters are held constant. The fiducial parameters are: line flux = $1\times10^{-17}$\,\fdens, observed FWHM$_\textrm{spec}=4$\,\AA\ (i.e. unresolved), line centred at the IFU centre, and FWHM$_\textrm{spat} = 0\arcsec$ (i.e. unresolved point source).}
\label{fig:completeness}
\end{figure*}

To evaluate the completeness of our emission line search we carry out comprehensive Monte Carlo simulations: inserting simulated lines into cubes with varied total flux, spectral FWHM$_\textrm{spec}$, spatial position, spatial extent FWHM$_\textrm{spat}$, and wavelength, and testing whether they are detected by our detection algorithm (Section~\ref{sec:results_lines}). Traditionally, these types of simulations are carried out by inserting simulated lines into real raw data and then running through the full reduction pipeline \citep{Fontana2010,Pentericci2014,DeBarros2017}, however, due to the complexity of the KMOS pipeline which constructs 3D cubes from 2D frames we instead create simulated cubes and add Gaussian noise drawn from an average noise cube for each cluster, mimicking completeness simulations traditionally done in imaging.

We create simulated flux cubes with a 3D Gaussian emission line with varied properties and add noise to each voxel drawn from a Gaussian distribution with mean zero and standard deviation $\sigma_{x,y,\lambda}$ for each cluster. The $\sigma_{x,y,\lambda}$ cubes are constructed by taking the RMS at every voxel of all the final sky-subtracted cubes which do not contain bright $z\simlt 2$ sources ($\sim10$ cubes per cluster). As each `empty' cube is expected to be pure noise, taking the RMS at each voxel across the cubes should give an estimate of the noise per voxel, $\sigma_{x,y,\lambda}$.

We calculate completeness as a function of flux, spatial offset from the IFU centre, spectral linewidth and spatial extent. For each simulation we vary the parameter of interest and wavelength, and fix the other three parameters. Our fiducial values for the parameters are: line flux = $1\times10^{-17}$\,\fdens, observed line FWHM$_\textrm{spec}=4$\,\AA\ (the spectral resolution, i.e. unresolved lines), line centred at the IFU centre, with source spatial extent FWHM$_\textrm{spat} = 0\arcsec$ (i.e. unresolved point source, the emission will have observed spatial extent with $\textrm{FWHM}_\textrm{spat,tot} = \sqrt{\textrm{FWHM}_\textrm{PSF}^2 + \textrm{FWHM}_\textrm{spat}^2}$). We draw 1000 realizations of an emission line with noise at every tested value of a parameter. The resulting completeness is the fraction of these simulated lines detected by our detection algorithm.

Our fiducial simulations assume \lya\ emission will be spatially unresolved. These assumptions are reasonable for the intrinsically UV faint LBGs we are observing \citep{Schmidt2016,Marchi2017}. Typical slit spectrograph observations of \lya\ emission centre slits on the UV continuum and use slit-widths $\sim0\farcs7$, thus in KLASS we are more complete to \lya\ emission that may be spatially extended and/or offset from the UV continuum.

Figure~\ref{fig:completeness} shows the results of our completeness simulations for all clusters. We reach $80\%$ completeness over the full wavelength range for lines $\simgt 5.7 \times 10^{-18}$\,\fdens, centred within $<0\farcs8$ of the IFU centre and with intrinsic line FWHM$_\textrm{spec} \simlt 250$\,\kms, assuming $z=8$ to calculate FWHM$_\textrm{spec}$ (median over all clusters). For wavelength ranges where the noise level is below the median across the whole spectrum, we reach $80\%$ completeness for $5\sigma$ lines $\simgt 3.2 \times 10^{-18}$\,\fdens, centred with $<0\farcs9$ of the IFU centre and with intrinsic line FWHM$_\textrm{spec} \simlt 550$\,\kms. The completeness is fairly flat for \lya\ spatial extent $\simlt0\farcs6$ (total extent $\simlt 0\farcs8$) demonstrating our good completeness for spatially extended \lya\ emission, with the normalisation of the completeness as a function of FWHM$_\textrm{spat}$ scaling with the completeness at a given total line flux.

\subsection{Flux and equivalent width limits}
\label{sec:results_flim}

\begin{figure*} 
\centering
\includegraphics[width=0.99\textwidth, clip=true]{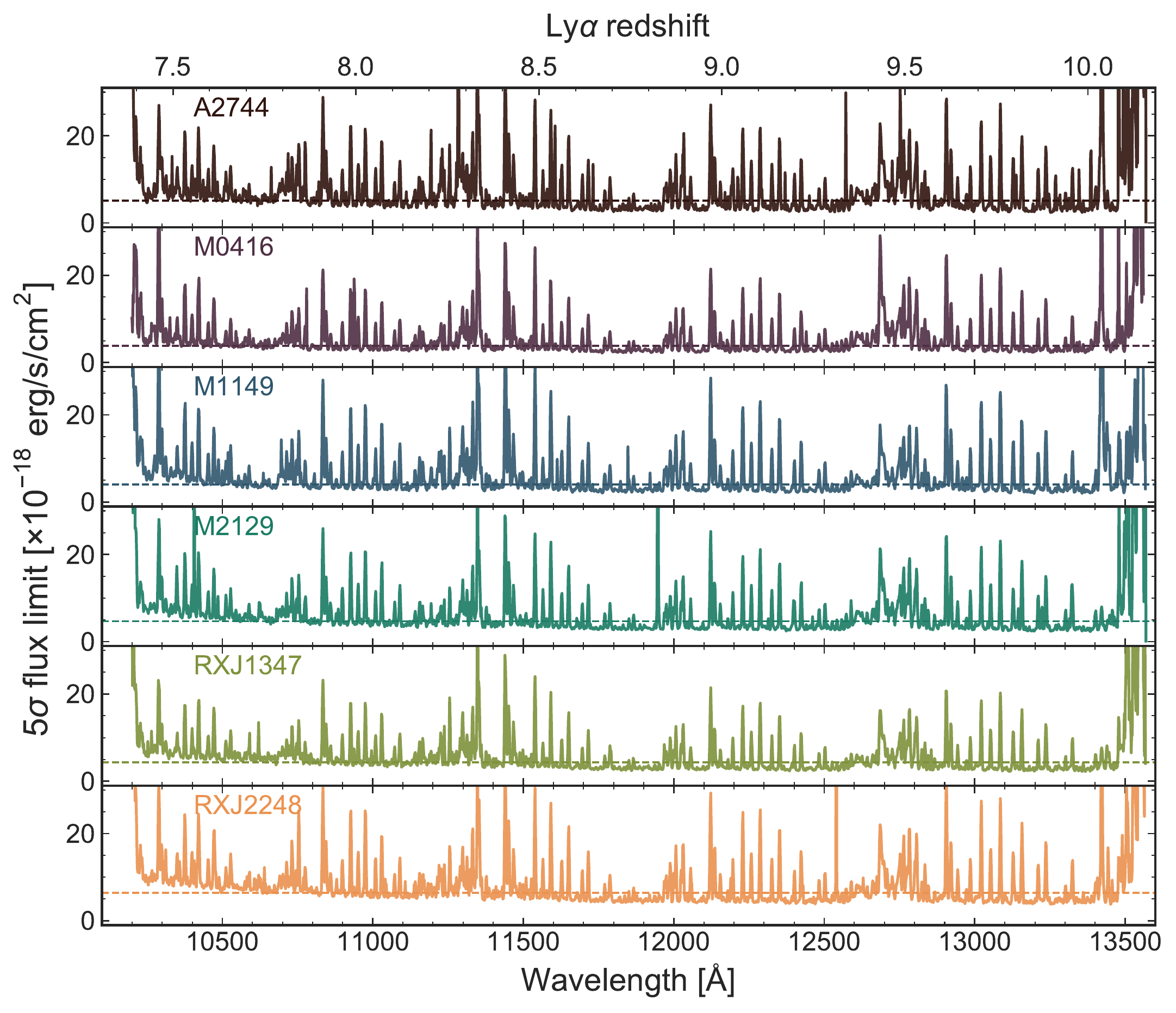}
\caption{Average $5\sigma$ flux limits for each cluster as a function of wavelength, assuming emission lines are spatially unresolved. We use the 1D RMS noise spectrum for each cluster as described in Section~\ref{sec:results_flim} to obtain the flux limits. Each plot corresponds to the median across all IFUs containing high-redshift candidates, for the different cluster targets. The dashed horizontal lines mark the median flux limit for each cluster.}
\label{fig:fluxlim}
\end{figure*}

\begin{figure*} 
\centering
\includegraphics[width=0.99\textwidth, clip=true]{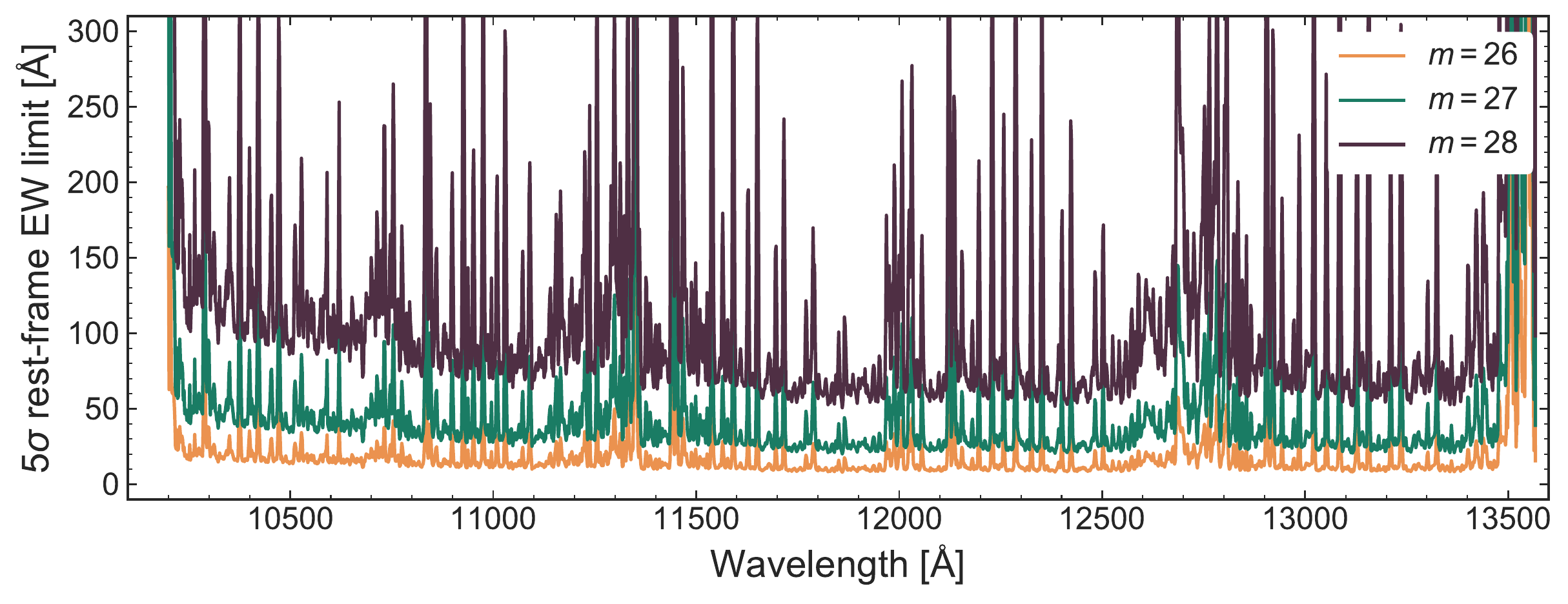}
\caption{$5\sigma$ rest-frame \lya\ EW limits in RXJ1347 as a function of wavelength, for 3 values of UV apparent magnitude $m$, assuming emission lines are spatially unresolved. We use the 5$\sigma$ flux limit for RXJ1347 shown in Figure~\ref{fig:fluxlim} and divide by the continuum flux and $(1+z_{\mathrm{Ly}\alpha})$ at each wavelength to obtain the EW limit.}
\label{fig:EWlim}
\end{figure*}

To calculate average flux limits for each cluster we take the average 3D noise spectrum for each cluster, $\sigma_{x,y,\lambda}$ (created by taking the RMS at every voxel across the $\sim10$ IFUs observing high redshift candidates in each cluster). We then create a 1D noise spectrum, $\sigma_\lambda$, by summing the average noise at each wavelength pixel in a circular aperture with radius $r=2\sigma_\textsc{psf}$ (where we use our median seeing $\mathrm{FWHM}_\textsc{psf}=0\farcs6$). At each wavelength pixel, $i$ the flux limit in \fdens\ is given by:
%
\BE \label{eqn:fluxlim}
f_{\textrm{lim},i} = 5 \times \frac{1}{1 - e^{-{\frac{r^2}{2\sigma_\textsc{psf}^2}}}} \sqrt{\frac{2FWHM_\mathrm{res}}{\Delta \lambda}}\sigma_i \times \Delta\lambda
\EE
Here we obtaining an estimate of the integrated noise for an emission line with observed $FWHM_\mathrm{res}=4$\,\AA\ or $\approx110$\,\kms\ (the instrumental resolution), and use a threshold integrated $S/N = 5$. The term in the denominator accounts for the fact that the apertures only capture a fraction of the flux. For $r=2\sigma_\textsc{psf}$ this results in a rescaling of 1.16. The spectral pixel width of KMOS YJ is $\Delta \lambda = 1.75$\,\AA. The above calculation assumes the emission is spatially and spectrally unresolved by KMOS, which is reasonable given the expectation that \lya\ emission from UV faint galaxies is likely to be more spatially compact and have lower linewidth than \lya\ from UV bright galaxies \citep[e.g.,][]{Schmidt2016,Marchi2017}. We note that the flux limit for wider lines can be estimated as $f_\textrm{lim} \propto \sqrt{FWHM/4\,\textrm{\AA}}$.


The $5\sigma$ flux limits for all clusters are shown in Figure~\ref{fig:fluxlim}. The median flux limit is $4.5 \times 10^{-18}$\,\fdens, and the range of medians for each cluster is $3.9 - 6.4 \times 10^{-18}$\,\fdens.

Rest-frame \lya\ equivalent widths are $W = (1+z)^{-1}f(\lambda)/f_\textrm{cont}$, where $z=\lambda/\lambda_\alpha - 1$ (with $\lambda_\alpha =1216$\,\AA), and we define the continuum flux:
\BE \label{eqn:fluxlim_fcont}
f_\textrm{cont}(m, z) = f_0 10^{-0.4m} \frac{c}{\lambda_\alpha^2(1+z)^2} \left(\frac{\lambda_\textsc{uv}}{\lambda_\alpha}\right)^{-\beta-2}
\EE
where $f_0 = 3.631 \times 10^{-20}$ erg s$^{-1}$ Hz$^{-1}$ cm$^{-2}$, $m$ is the apparent magnitude of the UV continuum, $c$ is the speed of light, $\lambda_\textsc{uv}$ is the rest-frame wavelength of the UV continuum (usually 1500\,\AA), and $\beta$ is the UV slope. We assume $\beta = -2$, consistent with $z\sim7$ observations \citep[e.g.,][]{Stanway2005,Blanc2011,Wilkins2011,Castellano2012,Bouwens2012,Bouwens2014}. We use the magnitude measured in \HST\ WFC3/F160W for the apparent magnitude (\texttt{automag}). Example EW limits for objects with a given apparent magnitude, using the RXJ1347 average flux limit, are plotted in Figure~\ref{fig:EWlim}.

\section{Reionization inference}
\label{sec:reionization}

In this section we describe the extension to the \citetalias{Mason2018a} Bayesian inference framework to include the full spectra, robustly including the uncertainties in redshift via the photometric redshift distribution (Section~\ref{sec:inference}), and marginalising over the linewidth of potential emission lines. Using the observations described above we now define a clear selection function for a photometrically-selected sample of LBGs within our survey (Section~\ref{sec:reionization_phot}), and perform the inference of the IGM neutral fraction using these data (Section~\ref{sec:reionization_infer}).

\subsection{Bayesian inference framework}
\label{sec:inference}

To use our observations to make inferences about the neutral hydrogen fraction at $z\sim8$ we use the method described by \citetalias{Mason2018a}. This forward-models the observed rest-frame \lya\ EW distribution as a function of the neutral fraction and galaxy UV magnitude, $p(W \,|\, \xHI, \MUV)$, using a combination of reionization simulations with realistic inhomogeneous IGM structure \citep{Mesinger2016a}, and empirical and semi-analytic models of galaxy properties. 

The models assume the observed $z\sim6$ \lya\ EW distribution is the `emitted' distribution (i.e. the distribution without IGM attenuation due to reionization) and use that to forward-model the observed distribution, including the impact of \lya\ velocity offsets. Here, as in \citetalias{Mason2018a}, we use the recent comprehensive $z\sim6$ \lya\ EW observations from \citet{DeBarros2017}. We use the public Evolution of 21cm Structure (EoS) suite of reionization simulations described by \citet{Mesinger2015,Mesinger2016a}\footnote{\url{http://homepage.sns.it/mesinger/EOS.html}} to generate \lya\ optical depths along millions of sightlines in simulated IGM cubes for a grid of volume-averaged $\xHI$ values. As the size of ionised regions during reionization is expected to be nearly independent of redshift at fixed $\xHI$ \citep[as there is little difference in the matter power spectrum from $z\sim7-11$,][]{McQuinn2007}, we use the same $z\sim7$ cubes as used by \citetalias{Mason2018a} rather than generating new $z\sim8$ cubes.

We refer the reader to \citetalias{Mason2018a} for more details of the forward-modelling approach. Here we describe the modifications we have made to our Bayesian inference to make use of the spectral coverage and sensitivity of our observations. We account for the incomplete redshift coverage and for the gravitational lensing magnification of the objects by the foreground clusters. We marginalise over a range of potential linewidths for the \lya\ emission lines. We also marginalise over the photometric redshift distribution for each galaxy, which we obtain from comprehensive photometry (Section~\ref{sec:reionization_phot}), to robustly account for uncertainties and degeneracies in redshift determination.

We want to obtain the posterior distribution for the neutral fraction: $p(\xHI \,|\, \{f\}, m, \mu)$ for each galaxy, where $\{f\}$ is an observed flux density spectrum as a function of wavelength, $m$ is the observed apparent UV magnitude, and $\mu$ is the magnification. A full derivation of the posterior is shown in Appendix~\ref{app:inference}, and we summarise it here.

Our inference framework calculates the likelihood of an emission line emitted at redshift $z_d$ with observed rest-frame EW, $=W$, being present in an observed flux density spectrum. To calculate this likelihood we must assume a lineshape for the observed emission line. In previous inferences by \citet{Treu2013,Pentericci2014} and \citet{Tilvi2014} treated emission lines as unresolved: lines were modelled as Dirac Delta functions, with all the flux contained in a single spectral pixel. However, motivated by recent observations of $z>6$ \lya\ emission with linewidth $\mathrm{FWHM}_\mathrm{spec} \sim200-450$\,\kms, several times greater than the instrumental resolution \citep{Ono2012,Finkelstein2013,Vanzella2011,Oesch2015,Zitrin2015a,Song2016a}, here we improve the method by including the effect of linewidth. 

The inference is quite sensitive to linewidth as at fixed EW a broader line will have lower S/N in our observations. By assuming unresolved emission lines the lower limits on the reionization `patchiness' parameter inferred by \citet{Treu2013,Pentericci2014} and \citet{Tilvi2014} will be slightly overestimated compared to a more realistic treatment of linewidth. We note that the $z\sim7$ neutral fraction inference by \citepalias{Mason2018a} used EW limits calculated assuming a range of realistic linewidths so their result does not need revision. We discuss the impact of linewidth in more detail in Appendix~\ref{app:inference_FWHM} but note that our results are robust for FWHM in a realistic range $~\sim100-400$\,\kms.

To modify our inference to account for linewidth, we assume Gaussian emission lines for simplicity so can write the model emission line flux density as a function of EW:
\BE \label{eqn:inference_linemod}
\begin{split}
f_\mathrm{mod}&(\lambda, W, m, z_d, \mathrm{FWHM}) = \\
&\frac{W f_{cont}(m,z_d)(1+z_d)}{\sqrt{2\pi}\sigma_\lambda} e^{-\frac{1}{2}\left(\frac{\lambda -\lambda_d}{\sigma_\lambda}\right)^2}
\end{split}
\EE
where $z_d = \lambda_d/\lambda_\alpha - 1$, with $\lambda_\alpha=1216$\,\AA, is the redshift of an emission line, $W$ is the rest-frame equivalent width of the emission line, $f_\mathrm{cont}$ is the flux density of the continuum calculated using Equation~\ref{eqn:fluxlim_fcont} using the observed continuum apparent magnitude $m$, and $\sigma_\lambda = FWHM/2.355$ is the spectral linewidth. 

The likelihood of observing a 1D flux density spectrum  $\{f\} = f(\lambda_i)$ for an individual galaxy (where $i$ is the wavelength pixel index), given our model where the true EW is drawn from the conditional probability distribution $p(W \,|\, \xHI, m, \mu, z_d)$ is:
\BE \label{eqn:inference_linelike}
\begin{split}
p&(\{f\} \,|\, \xHI, m, \mu, z_d, \mathrm{FWHM}) = \\
&\prod_i^N \int_0^\infty dW \, \biggl[\frac{1}{\sqrt{2\pi}\sigma_i}e^{-\frac{1}{2} \left(\frac{f_i - f_\mathrm{mod}(\lambda_i, W, m, z_d, \mathrm{FWHM})}{\sigma_i}\right)^2} \\
      	& \quad\quad\quad\quad\quad\quad \times p(W \,|\, \xHI, m, \mu, z_d) \biggr] 
\end{split}
\EE
where $\sigma_i$ is the uncertainty in flux density at wavelength pixel $i$ and there are a total of $N$ wavelength pixels in the spectrum. $p(W \,|\, \xHI, m, \mu, z_d)$ is the probability distribution for the observed rest-frame EW as a function of the neutral fraction, and galaxy properties -- UV apparent magnitude, magnification, and redshift. This PDF is obtained by convolving the $p(W \,|\, \xHI, \MUV)$ model outputs from \citetalias{Mason2018a} with the probability distribution for each galaxy's absolute UV magnitude, including errors on $m$ and $\mu$ (Equation~\ref{appeqn:inference_likeWMuv}). 

We note that the range of neutral fraction in the EoS simulations is $\xHI = 0.01 - 0.95$. In order to correctly calculate posteriors and confidence intervals we set the likelihood at $\xHI$ such that we expect to observe no \lya\ flux at all (in a fully neutral universe). I.e. $p(\{f\} \,|\, \xHI=1, m, \mu, z_d, \mathrm{FWHM}) = \prod_i^N \frac{1}{\sqrt{2\pi}\sigma_i}\exp{(-f_i^2/2\sigma_i^2)}$.

Given our relatively small sample size, we choose to restrict our inference to $z\sim8$, thus for ease of computation we evaluate $p(W \,|\, \xHI, m, \mu, z_d)$ at $z_d=8$; this has a negligible impact on the final likelihood. We keep $z_d$ free in the rest of the inference. This product of likelihoods over the wavelength range of the spectrum accounts for the wavelength sensitivity of our observations, i.e. high noise regions are weighted lower than low noise regions.

We also note that EW is independent of magnification. Therefore, our inferences should be quite robust to magnification, which enters only through the dependency on $\MUV$ of the assumed intrinsic EW distribution.

Using Bayes' theorem, the posterior distribution for $\xHI$, $z_d$ and FWHM is:
\BE \label{eqn:inference_post}
\begin{split}
p(\xHI, z_d, \mathrm{FWHM} \,|\, \{f\}, m, \mu) \propto{}& p(\{f\} \,|\, \xHI, m, \mu, z_d, \mathrm{FWHM}) \\
						& \times p(\xHI) \, p(z_d) \, p(\mathrm{FWHM})
\end{split}
\EE
We use a uniform prior on $\xHI$ between 0 and 1, $p(\xHI)$, and use the photometric redshift distribution for the prior $p(z_d)$. As we are only interested in the posterior probability of $\xHI$ we can marginalise over FWHM and $z_d$ for each galaxy. We use a log-normal prior on FWHM with mean depending on $\MUV$ derived through empirical relations and 0.3 dex width; we discuss our choice of FWHM priors in more detail in Appendix~\ref{app:inference_FWHM} but find our results to be negligibly changed if we had used a uniform prior spanning the range of observed \lya\ FWHM at $z>7$ ($\sim100-400$\,\kms). To account for the incomplete wavelength coverage, we use the fact that if the object has \lya\ outside of the KMOS wavelength range (covering $[z_\textrm{min}=7.2$, $z_\textrm{max}=10.1]$) we would measure a non-detection in our data. Thus the posterior for $\xHI$ from one galaxy is:
\BE \label{eqn:inference_postmarg}
\begin{split}
p(\xHI \,|\, \{f\}, m, \mu) &\propto \int_{z_\textrm{min}}^{z_\textrm{max}} dz_d \; p(\{f\} \,|\, \xHI, m, \mu, z_d) p(z_d)\\ 
	   &+ \prod_i^N \frac{1}{\sqrt{2\pi}\sigma_i}e^{-\frac{f_i^2}{2\sigma_i^2}}  \left( 1 - \int_{z_\textrm{min}}^{z_\textrm{max}} dz_d \; p(z_d) \right)
\end{split}
\EE
We assume all galaxies observed are independent, so that the final posterior is the product of the normalised posteriors (Equation~\ref{eqn:inference_postmarg}) for each object.

Using the photometric redshift distributions as a prior on the redshift allows us to incorporate the probability of each galaxy truly being at high redshift (rather than a low redshift contaminant) in a statistically rigorous way. In combining the posteriors in Equation~\ref{eqn:inference_postmarg} for each galaxy, the photometric redshift distribution weights the individual posteriors based on the probability of the source being within our redshift range. LBGs usually have degeneracies in their photometry which make it difficult to determine whether they are high redshift star-forming galaxies or mature $z\sim1-2$ galaxies. Thus with our method we are able to obtain reionization inferences from sources even when the photometric redshift distribution has multiple and/or broad peaks.

Whilst here we have carried out the inference at $z\sim8$ only, with larger samples, it will be possible to measure $\xHI(z)$ directly, for example by parametrising its evolution with redshift and inferring the values of its redshift-dependent parameters, or in a Markov Chain Monte Carlo exploration of IGM simulations to also infer relevant astrophysical parameters \citep{Greig2015,Greig2017}.

\subsection{Defining a selection function for a photometric sample}
\label{sec:reionization_phot}

\begin{figure} 
\centering
\includegraphics[width=0.49\textwidth, clip=true]{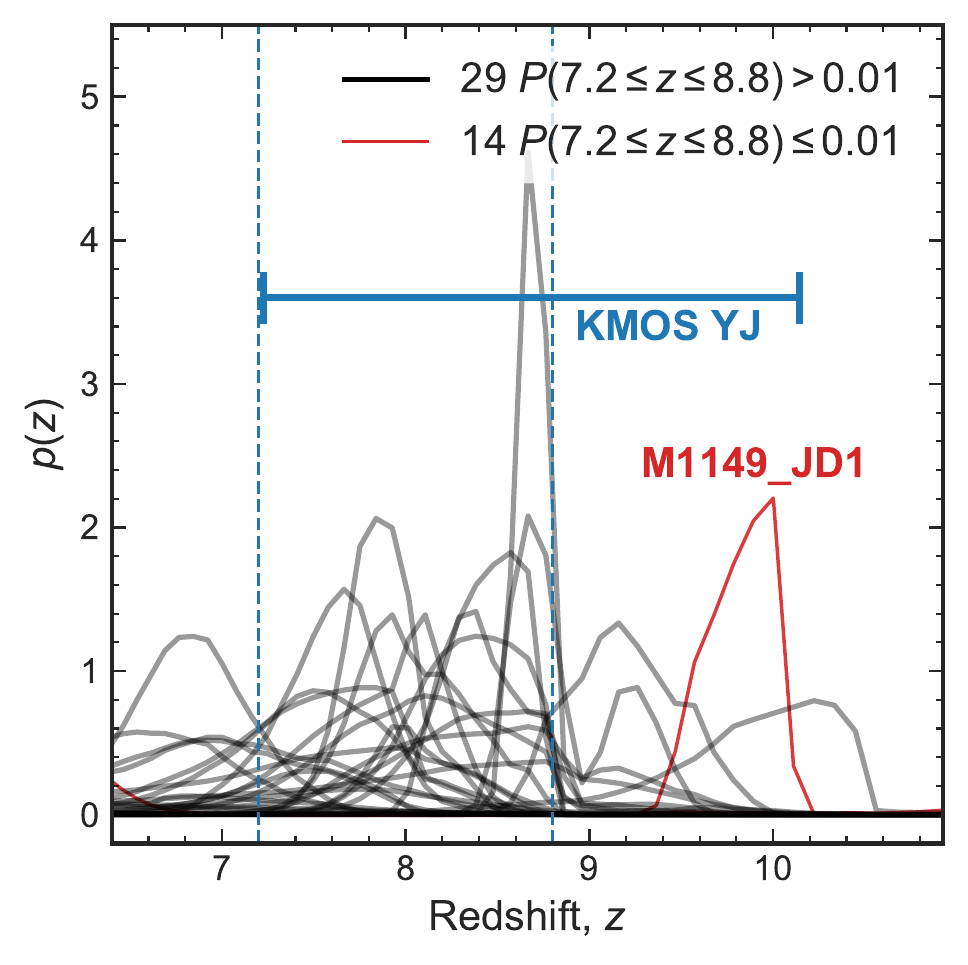}
\caption{Photometric redshift distributions centred on the KMOS observable range. We show the KMOS YJ range for \lya\ with the solid blue horizontal line. Black lines show the $p(z_\textrm{phot})$ for the 29 sources which have $>1\%$ probability of $7.2 \leq z_\textrm{phot} \leq 8.8$ (marked by blue dashed vertical lines) which we use for the inference. 14 sources have $P(7.2 \leq z_\textrm{phot} \leq 8.8) < 0.01$, including the galaxy M1149\_JD1, recently spectroscopically confirmed at $z=9.11$ with ALMA by \citet{Hashimoto2018}. In our photometric catalogue this galaxy is correctly found to be outside of our redshift range of interest (shown here as the red curve with $z_\textrm{phot}>9$), so we do not use it for our reionization analysis but discuss it in Section~\ref{sec:dis_JD1}. Note -- the remaining 13 objects have photometric redshift distributions outside of the range plotted here.}
\label{fig:pz}
\end{figure}

\begin{table*}
\centering
\caption{KLASS targets with $P(7.2 \leq z_\textrm{phot} \leq 8.8)$ solutions}
\label{tab:infersample}
\begin{tabular}[c]{lrrcccccc}
\hline
\hline 
Object ID$^\ast$ & R.A. & Dec. & $m_\mathrm{F160W}$ & $\mu$ & $M_\textsc{uv}^\dagger$ & $P(z_\textrm{phot})^\diamond$ & $f_\textrm{lim}^\ddagger\times 10^{-18}$ & $EW_{\textrm{Ly}\alpha}^{\ddagger,\star}$ \\
	   &    [deg]  &  [deg]  &           &       &                                   &                                        & [\fdens]                      & [\AA] \\
\hline 
A2744\_2036 & $3.596087$ & $-30.385836$ & $26.95\pm0.07$ & $2.4_{-0.5}^{+7.4}$ & $-19.27\pm0.89$ & 0.97 & $< 12.1$ & $<96$ \\
A2744\_2346 & $3.606460$ & $-30.380995$ & $26.78\pm0.06$ & $1.6_{-0.5}^{+0.8}$ & $-19.89\pm0.41$ & 1.00 & $< 10.6$ & $<70$ \\
A2744\_2345 & $3.606572$ & $-30.380932$ & $26.49\pm0.06$ & $1.6_{-0.5}^{+0.8}$ & $-20.19\pm0.41$ & 0.99 & $< 10.6$ & $<54$ \\
A2744\_2261 & $3.603996$ & $-30.382309$ & $27.29\pm0.10$ & $1.7_{-0.5}^{+1.1}$ & $-19.34\pm0.47$ & 0.79 & $< 10.6$ & $<113$ \\
A2744\_2503 & $3.588979$ & $-30.378668$ & $27.27\pm0.12$ & $2.2_{-0.7}^{+0.9}$ & $-19.04\pm0.39$ & 0.36 & $< 11.4$ & $<120$ \\
A2744\_2257 & $3.598123$ & $-30.382393$ & $28.62\pm0.18$ & $1.9_{-0.4}^{+0.8}$ & $-17.87\pm0.36$ & 0.54 & $< 10.7$ & $<392$ \\
A2744\_20236 & $3.572523$ & $-30.413267$ & $28.61\pm0.24$ & $1.8_{-0.5}^{+1.0}$ & $-17.94\pm0.48$ & 0.42 & $< 9.5$ & $<342$ \\
A2744\_1040 & $3.592505$ & $-30.401482$ & $27.52\pm0.15$ & $14.2_{-6.3}^{+11.2}$ & $-16.79\pm0.65$ & 0.04 & $< 9.7$ & $<129$ \\
A2744\_2248$^{\ast\ast}$ & $3.603863$ & $-30.382261$ & $26.57\pm0.07$ & $1.7_{-0.5}^{+1.1}$ & $-20.06\pm0.47$ & 0.96 & $< 10.6$ & $<58$ \\
M0416\_99$^{\ast\ast\ast}$ & $64.039162$ & $-24.093182$ & $26.28\pm0.05$ & $1.5_{-0.3}^{+0.5}$ & $-20.49\pm0.30$ & 0.78 & $< 3.4$ & $<14$ \\
M0416\_286 & $64.037567$ & $-24.088116$ & $28.20\pm0.17$ & $1.9_{-0.5}^{+0.3}$ & $-18.29\pm0.31$ & 0.66 & $< 3.6$ & $<89$ \\
M0416\_743 & $64.048058$ & $-24.081427$ & $26.56\pm0.06$ & $1.7_{-0.2}^{+0.3}$ & $-20.07\pm0.19$ & 0.07 & $< 3.1$ & $<17$ \\
M0416\_1956 & $64.060333$ & $-24.064962$ & $28.16\pm0.16$ & $1.9_{-0.6}^{+0.2}$ & $-18.33\pm0.28$ & 0.91 & $< 3.0$ & $<72$ \\
M0416\_1997 & $64.049583$ & $-24.064596$ & $27.56\pm0.17$ & $6.3_{-1.5}^{+39.3}$ & $-17.64\pm1.23$ & 0.90 & $< 2.9$ & $<40$ \\
M0416\_22746 & $64.046509$ & $-24.061630$ & $27.77\pm0.23$ & $8.1_{-3.0}^{+4.3}$ & $-17.15\pm0.53$ & 0.62 & $< 2.9$ & $<49$ \\
M1149\_23695 & $177.382996$ & $22.412041$ & $28.11\pm0.14$ & $3.6_{-2.1}^{+0.7}$ & $-17.69\pm0.58$ & 0.77 & $< 4.2$ & $<97$ \\
M1149\_3343 & $177.392715$ & $22.384718$ & $28.64\pm0.28$ & $1.7_{-0.5}^{+0.4}$ & $-17.96\pm0.42$ & 0.04 & $< 5.3$ & $<201$ \\
M1149\_1428 & $177.412216$ & $22.394894$ & $28.34\pm0.17$ & $7.5_{-2.8}^{+0.9}$ & $-16.67\pm0.36$ & 0.25 & $< 3.1$ & $<87$ \\
M1149\_945 & $177.412079$ & $22.389055$ & $27.92\pm0.13$ & $9.2_{-3.2}^{+14.4}$ & $-16.87\pm0.76$ & 0.16 & $< 3.3$ & $<63$ \\
M2129\_2633 & $322.345232$ & $-7.671373$ & $25.65\pm0.12$ & $1.6_{-0.1}^{+0.1}$ & $-21.06\pm0.13$ & 0.20 & $< 3.4$ & $<8$ \\
M2129\_2661 & $322.350848$ & $-7.675239$ & $26.38\pm0.17$ & $1.7_{-0.0}^{+0.0}$ & $-20.25\pm0.17$ & 0.07 & $< 3.4$ & $<15$ \\
M2129\_1556 & $322.344535$ & $-7.688473$ & $27.53\pm0.26$ & $4.2_{-0.2}^{+0.2}$ & $-18.11\pm0.27$ & 0.01 & $< 3.4$ & $<45$ \\
RXJ1347\_1831 & $206.896270$ & $-11.742338$ & $26.30\pm0.26$ & $9.2_{-0.4}^{+0.4}$ & $-18.49\pm0.26$ & 0.06 & $< 3.3$ & $<14$ \\
RXJ1347\_656 & $206.891246$ & $-11.752607$ & $26.43\pm0.24$ & $20.4_{-1.2}^{+1.6}$ & $-17.49\pm0.25$ & 0.72 & $< 3.7$ & $<18$ \\
RXJ1347\_101 & $206.880973$ & $-11.769816$ & $25.16\pm0.15$ & $43.9_{-5.4}^{+10.2}$ & $-17.92\pm0.26$ & 0.20 & $< 3.6$ & $<5$ \\
RXJ1347\_1368 & $206.893076$ & $-11.760230$ & $27.92\pm0.43$ & $16.6_{-1.1}^{+1.1}$ & $-16.22\pm0.43$ & 0.34 & $< 3.1$ & $<60$ \\
RXJ1347\_1280 & $206.896921$ & $-11.763833$ & $27.28\pm0.28$ & $4.8_{-0.5}^{+0.7}$ & $-18.22\pm0.31$ & 0.03 & $< 2.8$ & $<29$ \\
RXJ2248\_1006 & $342.208379$ & $-44.537520$ & $25.83\pm0.17$ & $1.6_{-0.4}^{+0.4}$ & $-20.88\pm0.32$ & 0.92 & $< 4.6$ & $<13$ \\
RXJ2248\_2086 & $342.179829$ & $-44.525664$ & $26.88\pm0.13$ & $41.0_{-25.5}^{+72.3}$ & $-16.28\pm1.09$ & 0.48 & $< 3.9$ & $<28$ \\
\hline
\multicolumn{9}{p{0.99\textwidth}}{\textsc{Note.} -- $^\ast$ IDs for A2744, M0416 and M1149 match the ASTRODEEP catalogue IDs \citep{Merlin2016,DiCriscienzo2017}. $^\dagger$ These listed intrinsic magnitudes are calculated using $z=8$ and the listed magnifications and errors. $^\diamond$ This is the photometric redshift from EAzY integrated between $z=7.2$ and $z=8.8$, i.e. the total probability of the object to have. $^\ddagger$ Flux and EW limits are $5\sigma$. $^\star$ All EW are rest-frame. We stress that the EW limits only hold if the \lya\ is actually in the KMOS range, which has probability given by $P(7.2 \leq z_\textrm{phot} \leq 8.8)$. $^{\ast\ast}$ This object was spectroscopically confirmed by \citet{Laporte2017} at $z=8.38$. $^{\ast\ast\ast}$ This object was spectroscopically confirmed by \citet{Tamura2018} at $z=8.31$.}
\end{tabular}
\end{table*}


To make accurate inferences for reionization it is important to have uniform and well-understood target selection functions for the sources we use. At the time of target selection for KLASS not all deep HFF data were available, nor were sophisticated ICL removal techniques developed \citep[e.g.,][]{Merlin2016,Morishita2017a,Livermore2017}. This led to heterogeneous target selections. However, for this analysis we now use the most up-to-date photometry available to create a sub-sample for analysis with a homogeneous selection function. We demonstrate in Appendix~\ref{app:phot} that this sub-sample is not a biased selection from the final parent catalogues.

Deep, multi-band \HST, Spitzer-IRAC and HAWK-I photometry is now available for all our targets through the CLASH, SURFSUP, and HFF programs \citep{Postman2012,Bradac2014,Huang2016,Lotz2017}. For A2744, M0416 and M1149 we used the ASTRODEEP photometric catalogue which removed foreground intra-cluster light \citep{Castellano2016b,Merlin2016,DiCriscienzo2017}. For M2129, RXJ1347 and RXJ2248 we created our own catalogues based on the ASTRODEEP methodology (M. Brada\v{c} et al., in prep). Of the 56 high-redshift candidate targets we assigned to KMOS IFUs, 46 have matches in these final deep catalogues (including the 3 images of a $z=6.11$ multiply-imaged system in RXJ2248).

To determine why 10 targets had no match in the final photometric catalogues we examined our target selection catalogues. We used preliminary versions of the ASTRODEEP catalogues for A2744, M0416 and M1149 in our initial selection, so all the objects targeted in A2744 and M0416 have matches in the final catalogues. 3 targets do not appear in the final M1149 catalogue, these objects were never in the preliminary ASTRODEEP catalogue but were selected from alternative preliminary HFF catalogues. 3 targets from M2129, 3 targets from RXJ1347 and 1 target from RXJ2248 have no matches in the final catalogues, which was expected as they were selected from an ensemble of preliminary photometric catalogues with shallower photometry, and narrower wavelength coverage compared to our final catalogues. 3 of the unmatched objects were Category 1 targets. These missing targets were likely faint in the initial photometry and so turn out to be spurious in deep photometry.

Photometric redshift distributions were obtained from the final catalogues with the EAzY code \citep{Brammer2008}. We perform the EAzY fit to the entire photometric dataset, and obtain photometric redshift posteriors without the magnitude prior (which weights bright objects to lower redshifts based on observations of field galaxies and may be inappropriate for our lensed sources). As described in Section~\ref{sec:inference} our inference framework uses the full photometric redshift distribution thus we can robustly use all objects with non-zero probability of being in our redshift range of interest for our inferences.

Taking the 43 high redshift KMOS targets matched in the catalogues (excluding the three images of the $z=6.11$ galaxy described in Appendix~\ref{app:CIV}) we then use the photometric redshift distributions to select objects which could be in the KMOS YJ range ($ 7.2 < z_\textrm{phot} < 10.1$). We calculate $P(7.2 < z_\textrm{phot} < 10.1) = \int_{7.2}^{10.1} p(z_\textrm{phot}) dz_\textrm{phot}$ using the normalised EAzY photometric redshift distribution for each object to find the total probability of the object being within that redshift range. We select 30 objects with $P(7.2 < z_\textrm{phot} < 10.1) > 0.01$ (though the majority have a much higher probability of being in that redshift range). The photometric redshift distributions of these objects within the KMOS YJ range are plotted in Figure~\ref{fig:pz}. 

We examined the final deep photometry of the 13 objects which dropped out of the KMOS YJ range in this selection, which include 6 Category 1 targets. As expected, the selection of these objects shifts to lower redshifts now the full photometry is available. The majority of them have detections in the bluest bands which would negate a $z>7$ Lyman Break, and several are clearly $z\sim1$ passive galaxies when the IRAC bands are included.

Due to the relatively small sample size, we choose to perform our inference at $z\sim8$, so we select only objects with some probability to have $7.2 < z_\textrm{phot} < 8.8$. We calculate $P(7.2 < z_\textrm{phot} < 8.8) = \int_{7.2}^{8.8} p(z_\textrm{phot}) dz_\textrm{phot}$. We select 29 objects with $>1\%$ probability of being within this redshift range (21 have $>10\%$ probability, and 13 $> 60\%$ probability). One object has $z_\textrm{phot} > 9$ and is excluded from our inference. This is M1149\_JD1, recently spectroscopically confirmed at $z=9.11$ by \citet{Hashimoto2018} in ALMA, who also show a tentative \lya\ detection from X-shooter. As this galaxy's photometric redshift distribution clearly puts it at $z>9$ we do not include it in our $z\sim8$ reionization inferences. Its $p(z_\textrm{phot})$ can be seen in Figure~\ref{fig:pz} (red line) and we discuss our observations of it in Section~\ref{sec:dis_JD1}.

Our inference uses the full $p(z_\textrm{phot})$ distribution, to robustly account for any probability of an object being a lower redshift contaminant. The median and standard deviation of best-fit photometric redshifts over this range for the sub-sample of 29 objects is $z_\textrm{phot} = 7.9\pm0.6$. These objects and their observed properties, including $P(7.2 < z_\textrm{phot} < 8.8)$ are listed in Table~\ref{tab:infersample}. We demonstrate that this sub-sample is not a biased sample of the final photometric catalogues in Appendix~\ref{app:phot}.

We also cross-checked our Table~\ref{tab:infersample} with publicly available spectroscopic catalogues from ground-based follow-up at optical wavelengths for clusters A2744 \citep{Mahler2018}, M0416 \citep{Balestra2016,Caminha2017}, M1149 \citep{Grillo2016}, M2129 \citep{Monna2017} and RXJ2248 \citep{Karman2015,Karman2016}. We found no matches in those catalogues for any of the objects in Table~\ref{tab:infersample}. Non-detections of these objects in optical spectroscopy lends credence to their selection as $z>7$ candidates. 

Two objects have been spectroscopically confirmed at $7.2 < z < 8.8$ by groups. A2744\_2248 (a.k.a. A2744\_YD4) was confirmed at $z=8.38$ via [\OIII]88$\mu$m emission in ALMA, and a tentative \lya\ emission line was also reported with line flux $(1.82\pm0.64) \times10^{-18}$\,\fdens\ and $EW=10.7\pm2.7$ \citep{Laporte2017}, which is well below our limit for that object. As discussed in Section~\ref{sec:reionization_infer} we find that treating the object as a detection in our inference has a negligible impact on our inferred limits on the neutral fraction. M0416\_99 (a.k.a M0416\_Y1) was also confirmed via [\OIII]88$\mu$m emission in ALMA observations at $z=8.31$ by \citet{Tamura2018}. They also observed the object with X-shooter and found no rest-frame UV emission lines, with a $5\sigma$ \lya\ flux limit of $<8.0\times10^-18$\,\fdens\ (if the line if offset by up to 250\,\kms). Our KMOS flux median limits are of a comparable depth ($<3.4\times10^-18$\,\fdens).

We obtain magnification estimates for each object using the publicly available HFF lens models\footnote{\url{https://archive.stsci.edu/prepds/frontier/lensmodels/}}. We take the best-fit magnifications from the most recent versions of all available lens models for each object, drop the highest and lowest magnifications to produce an approximate $1\sigma$ range of estimated magnifications, $\{ \mu \}$. We list the median magnification from this sub-sample, and the upper and lower bounds in Table~\ref{tab:infersample}. For the inference, we assume magnifications are log-normally distributed with mean given by the median $\log_{10}\{ \mu \}$ and standard deviation given by half the range of $\log_{10}\{ \mu \}$, which is a reasonable fit to the distribution of magnifications from the models. For M2129 and RXJ1347, the only non-HFF clusters, we use the magnification distribution from the Brada{\v c} group lens models \citep{Huang2016a,Hoag2019} and obtain mean and standard deviation log magnifications. As discussed in Section~\ref{sec:inference} by using the EW in our inference, which is independent of magnification (as opposed to flux), our results are quite robust to magnification uncertainties.

We calculate flux and \lya\ EW limits for individual objects as in Section~\ref{sec:results_flim}, using Equations~\ref{eqn:fluxlim} and \ref{eqn:fluxlim_fcont}. The median intrinsic UV absolute magnitude (i.e., corrected for magnification) of the sample is $\MUV = -18.2$. The median observed flux $5\sigma$ upper limit in this sub-sample is $< 3.6\times10^{-18}$\,\fdens, and the median rest-frame \lya\ EW $5\sigma$ upper limit is $< 58$\,\AA.

\subsection{Inference on the IGM neutral fraction}
\label{sec:reionization_infer}

We use 1D spectra and uncertainties as a function of wavelength for the 29 objects described above to infer the IGM neutral fraction at $z\sim8$ using Equation~\ref{eqn:inference_postmarg} to calculate the posterior distribution of $\xHI$. 

We obtain the flux density spectra using the cubes for each object, extracting flux and noise in a circular aperture with $r=2\sigma_{psf}$, and apply a rescaling to both to account for the incomplete recovery of flux in the aperture, and a constant rescaling to the noise spectrum to ensure the S/N distribution of pixels in each spectrum is a Normal distribution.

\begin{figure}
\centering 
\includegraphics[width=0.49\textwidth, clip=true]{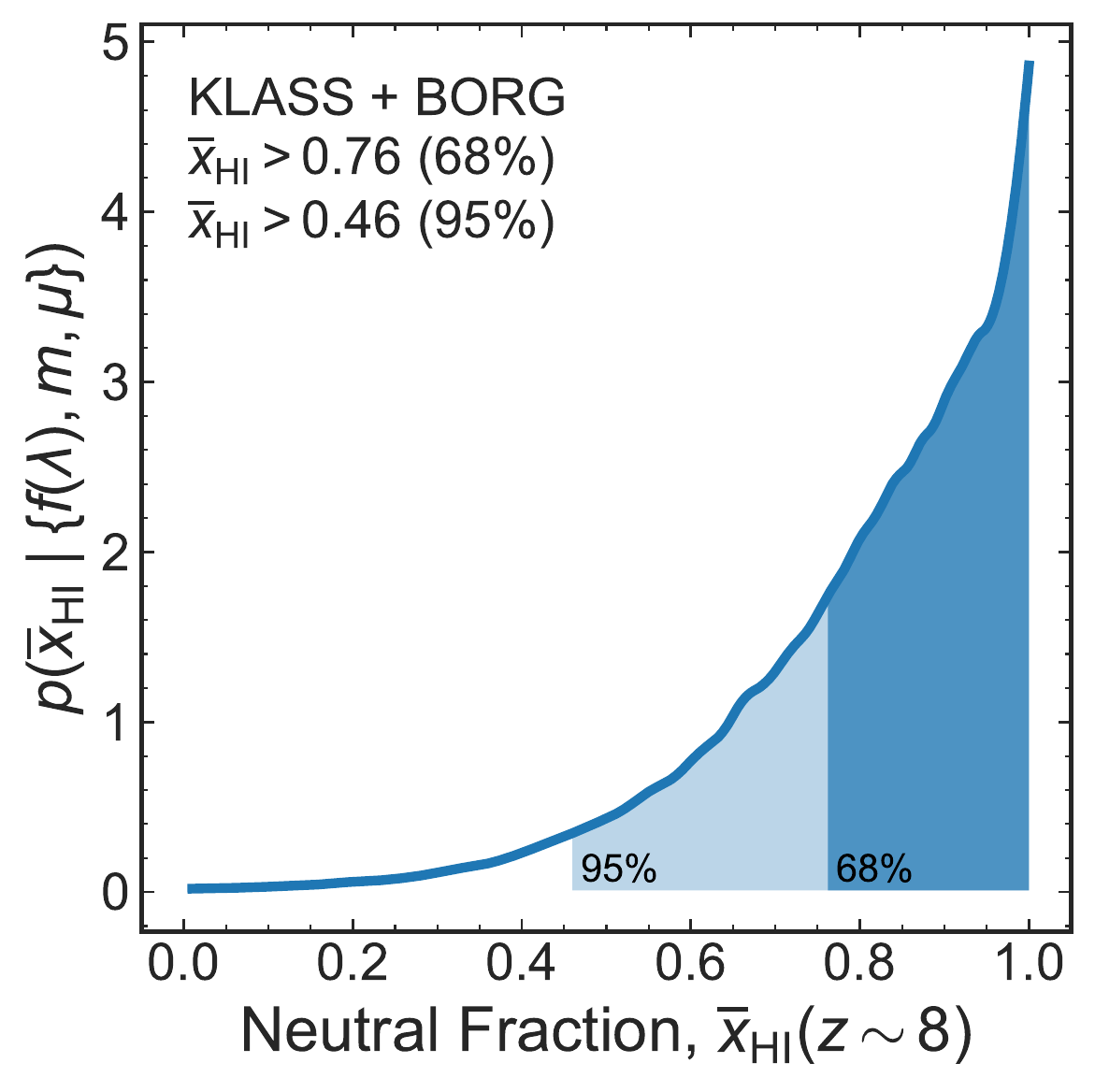}
\caption{Posterior probability distribution for the IGM neutral fraction $\xHI$ at $z\sim8$ obtained using Equation~\ref{eqn:inference_postmarg} and the EW spectra from the KLASS sample described in Section~\ref{sec:reionization_phot} and the BoRG sample described by \citet{Treu2013}. The blue line and shaded regions show the posterior from the combined datasets, and its 68\% and 95\% confidence regions (the darkest region is the 68\% confidence range).}
\label{fig:xHI}
\end{figure}

In Figure~\ref{fig:xHI} we plot the posterior distribution for $\xHI$ obtained using our observations of the 29 $z\sim8$ KLASS targets, as well as Keck/MOSFIRE observations of 8 $z\sim8$ LBGs from the Brightest of Reionizing Galaxies survey \citep[BoRG,][]{Trenti2011,Bradley2012,Schmidt2014a} described by \citet{Treu2013}. Using the BoRG sample allows us to cover a broader range in intrinsic magnitudes spanning opposite ends of the galaxy UV luminosity function: the IGM attenuation of \lya\ from UV bright and UV faint galaxies is expected to be different due to differing \lya\ escape paths through their interstellar and circumgalactic media \citep[e.g.,][]{Stark2010,Stark2017,Mason2018b}.

These two sets of independent observations, both indicate a predominantly neutral IGM at $z\sim8$. The BoRG data alone produce a lower limit of $\xHI > 0.34$ (68\%) and for the KLASS data alone $\xHI > 0.76$ (68\%). Lower limits from the combined dataset are $\xHI > 0.76$ (68\%) and $\xHI > 0.46$ (95\%).

By exploiting gravitational lensing, the KLASS sample sets much lower limits on the \lya\ EW for intrinsically UV faint galaxies \citepalias[which produce the strongest constraints on reionization's mid-stages,][]{Mason2018a} than is possible in blank fields. Our new KLASS sample also demonstrates how increasing the number of sources for the inference produces much tighter constraints on the IGM neutral fraction compared to the 8 BoRG sources.

To test whether the inclusion of objects with candidate \lya\ emission in GLASS data biased our sample, we tested the inference with and without including the Category 1 targets (which were specifically targeted in KLASS because they had candidate \lya\ emission in the \HST\ data). We found no significant difference in the posteriors. We also tested the inference with and without including the $z=8.38$ marginal detection of \lya\ in object A2744\_2248 by \citet{Laporte2017} with spectroscopic confirmation from \OIII\ emission in ALMA observations. We use the EW reported by \citet{Laporte2017} $W=10.7\pm2.7$\,\AA, which is well below our $5\sigma$ limit for that object ($<53$\,\AA). Despite the potential detection, the posterior distribution for this single object strongly favours a mostly neutral IGM due to its very low EW and low significance. We did our inference using both our KMOS spectra and the \citet{Laporte2017} measurement for this object and found it to have a negligible impact on our final posterior (changing the inferred limit by only $\Delta \xHI \sim 0.01$), demonstrating that deep limits on non-detections have a lot of power in our inferences. Our quoted posterior limits include the object as a non-detection.

\section{Discussion}
\label{sec:dis}

We discuss our new lower limit on the neutral fraction and the implications for the timeline of reionization in Section~\ref{sec:dis_reionization}, and show it favours reionization driven by UV faint galaxies with a low ionizing photon escape fraction. In Section~\ref{sec:dis_JD1} we discuss the recent tentative detection of \lya\ at $z=9.11$ by \citep{Hashimoto2018} and show it is not inconsistent with our results. In Section~\ref{sec:dis_otherUV} we discuss our EW limits on NV and \CIV\ emission. Finally, in Section~\ref{sec:dis_KMOS} we present a comparison of the KMOS ETC and our achieved S/N for background-limited observations.

\subsection{The timeline of reionization}
\label{sec:dis_reionization}

\begin{figure} 
\centering
\includegraphics[width=0.49\textwidth, clip=true]{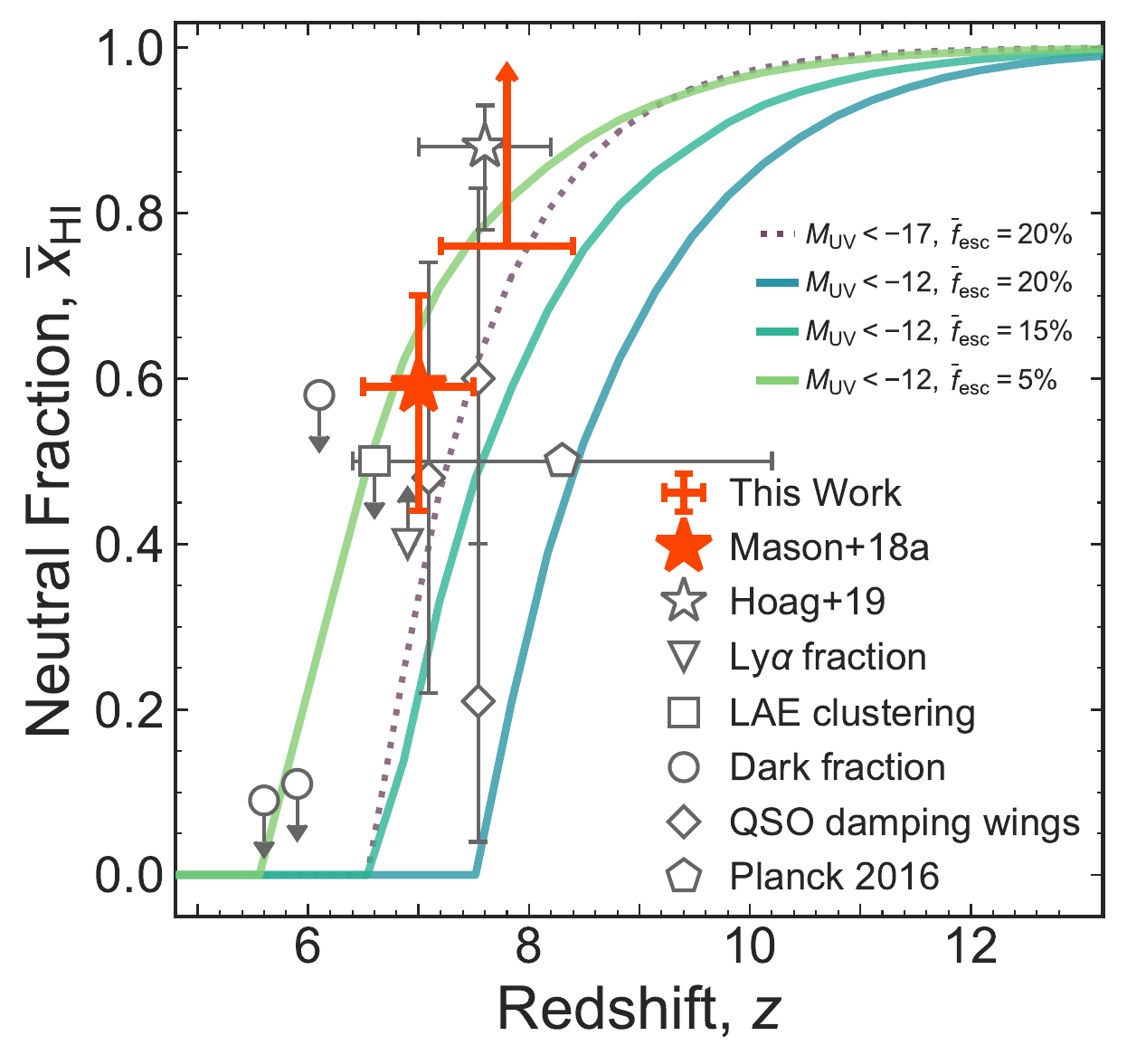}
\caption{The redshift evolution of the volume average neutral hydrogen fraction of the IGM. Our new lower limit is shown in orange, with the horizontal errorbar at the 68\% confidence level. We also plot measurements derived from observations of: the evolving \lya\ EW distribution at $z\sim7$ \citepalias[orange filled star][]{Mason2018a} previous estimates from the fraction of LBGs emitting \lya\ \citep[open black star,][]{Mesinger2015}; the clustering of \lya\ emitting galaxies \citep[square,][]{Ouchi2010,Sobacchi2015}; \lya\ and Ly$\beta$ forest dark fraction \citep[circle - 68\% limits,][]{McGreer2014}; and QSO damping wings \citep[diamond,][]{Davies2018,Greig2018b}. We offset the constraints at $z\sim7$ by $\Delta z=0.1$ for clarity. We also plot the \citet{PlanckCollaboration2016} redshift range of instantaneous reionization (black pentagon). We show median model reionization histories derived from the \citet{Mason2015} UV luminosity function models as coloured lines. We plot models obtained from integrating the luminosity function down to two magnitude limits -- $M_\textsc{uv} =-17$ (purple dashed line) and $M_\textsc{uv} = -12$ (darkest blue solid line) and drawing from uniform distributions for the ionizing photon escape fraction $10-30$\% ($\langle f_\textrm{esc} \rangle = 20\%$) and clumping factor $C=1-6$, and log-normal distribution for the ionizing efficiency $\xi_\textrm{ion}$ with mean $25.2$ and standard deviation $0.15$\,dex. Comparing reionization histories with ionizing escape fraction drawn from a uniform distribution $1-10\%$ (light green, $\langle f_\textrm{esc} \rangle \approx 5\%$) and $10-20\%$ (medium teal, $\langle f_\textrm{esc} \rangle = 15\%$), integrating LFs down to $M_\textsc{uv} = -12$ in both cases and using the same distribution for the clumping factor and $\xi_\textrm{ion}$ as above.}
\label{fig:xHI_history}
\end{figure}

\begin{figure} 
\centering
\includegraphics[width=0.49\textwidth, clip=true]{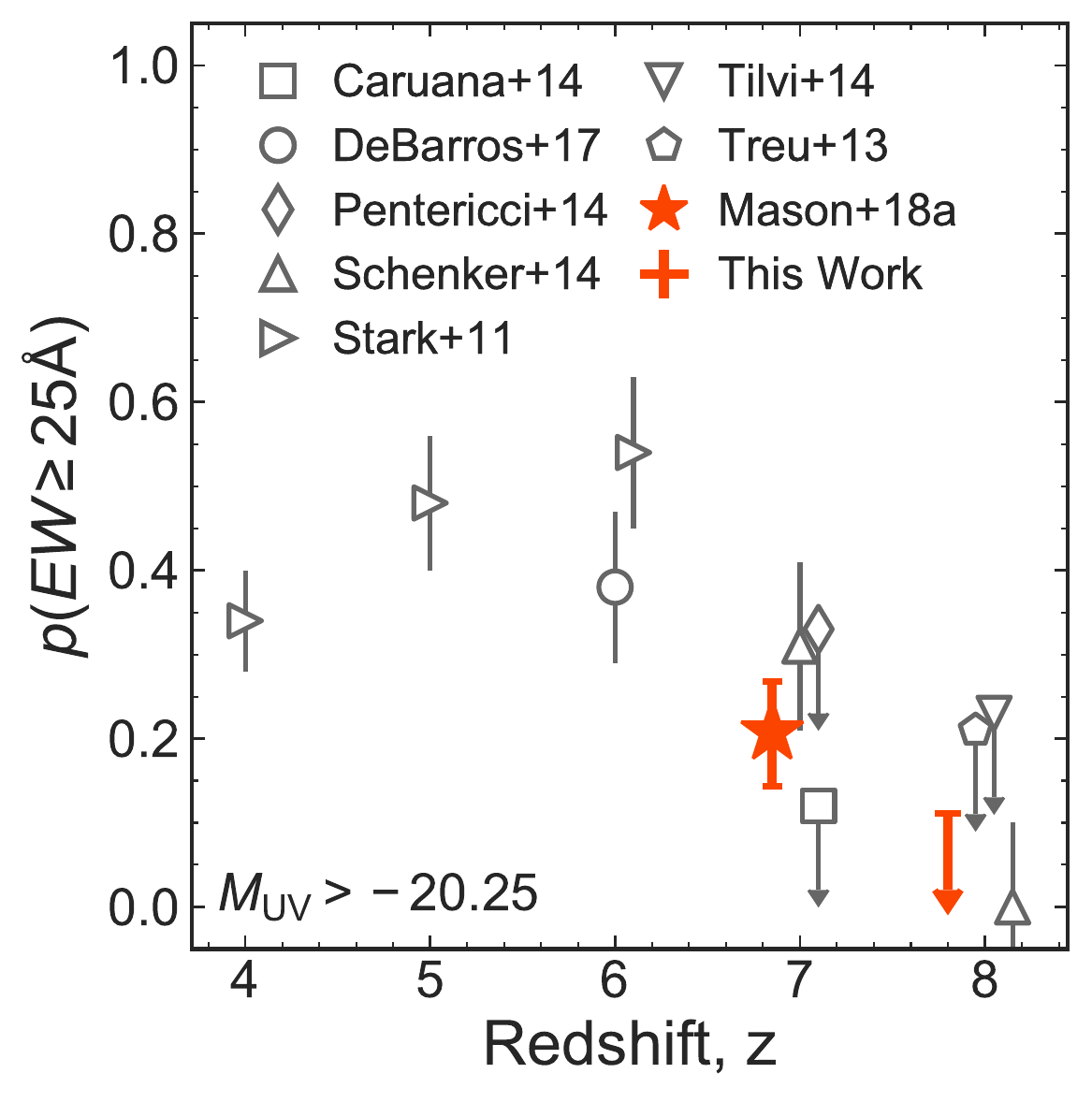}
\caption{The redshift evolution of the `\lya\ fraction' for UV faint galaxies, the fraction of LBGs observed with \lya\ EW $\geq25$\,\AA. We plot literature measurements from \citet{Stark2011,Pentericci2014,Treu2013,Tilvi2014,Schenker2014,Caruana2014} and \citet{DeBarros2017}. We add small offsets in redshift for measurements at the same redshifts to ease the display of the data. We also plot the predicted \lya\ fraction from \citetalias{Mason2018a} calculating $p(W > 25\AA \,|\, \xHI, \MUV)$ using $\MUV = -20$ galaxies and the neutral fraction constraint $\xHI = 0.59_{-0.15}^{+0.11}$ ($16-84\%$ confidence intervals) as the orange star. We plot the upper limits recovered in this paper as orange lines, with the solid line showing our 68\% confidence limit, and the dotted line extending to the 95\% confidence limit. We calculate $p(W > 25\AA \,|\, \xHI > 0.76, \MUV)$ again using $\MUV = -20$. Our constraint is consistent with literature values at the same redshift.}
\label{fig:Lya_frac}
\end{figure}

We plot our new limit on the reionization timeline in Figure~\ref{fig:xHI_history}. We also plot other statistically robust constraints from \citet{Ouchi2010,McGreer2014,Sobacchi2015,Mesinger2015,Davies2018,Greig2018b,Mason2018a} and the \citet{PlanckCollaboration2016}. While no increase in $\xHI$ compared to the \citet{Mason2018a} constraint is statistically possible within the 95\% confidence interval, our new limit, combined with the other recent $\xHI$ statistical measurements at $z\sim7$, and other estimates \citep[e.g.,][]{Caruana2014,Zheng2017a}, provides increasing evidence for the bulk of hydrogen reionization occurring $z\sim6-8$ \citep{Greig2016,Banados2017,Davies2018,Mason2018a}, late in the \citet{PlanckCollaboration2016} confidence range.

Accurate measurements of the reionization timeline can help constrain properties of early galaxies. In Figure~\ref{fig:xHI_history} we show model reionization histories obtained from integrating the \citet{Mason2015} UV luminosity functions, varying the typical reionization parameters: the minimum UV luminosity of galaxies, and the average ionizing photon escape fraction. We see that late reionization is most consistent with either a high minimum UV luminosity of galaxies ($\MUV < -17$) and moderate escape fraction ($\langle f_\textrm{esc} \rangle = 20\%$), or with including ultra-faint galaxies $\MUV < -12$) with low escape fractions ($\langle f_\textrm{esc} \rangle \simlt 15\%$). 

There are many degeneracies between these reionization parameters, and certainly the escape fraction is unlikely to be constant for all galaxies at all times \citep{Trebitsch2017}, but non-detections of high-redshift GRB host galaxies, and observations of lensed high-redshift galaxies, and local dwarfs, indicate galaxies fainter than $M_\textsc{uv} = -17$ likely exist at $z\sim8$ \citep[e.g.,][]{Kistler2009,Tanvir2012,Trenti2012a,Alavi2014,Weisz2017,Livermore2017,Bouwens2017,Ishigaki2018}. If ultra-faint galaxies do contribute significantly to reionization our result suggests reionization can be completed with low escape fractions, consistent with low redshift estimates of the average escape fraction \citep{Marchi2017,Rutkowski2017,Naidu2018,Steidel2018}.

For comparison with previous high redshift \lya\ spectroscopic surveys we plot the so-called `\lya\ fraction', the fraction of LBGs emitting \lya\ with EW $\geq25$\,\AA\ in Figure~\ref{fig:Lya_frac}. We compare our new upper limits on the \lya\ fraction with literature measurements from \citet{Stark2011,Pentericci2011,Treu2013,Tilvi2014,Schenker2014,Caruana2014} and \citet{DeBarros2017}. We also plot the predicted \lya\ fraction from \citetalias{Mason2018a}. Using the \citetalias{Mason2018a} model EW distributions $p(W \,|\, \xHI, \MUV)$ we can calculate the \lya\ fraction as the probability of EW $\geq 25$\,\AA\ given our constraint on the neutral fraction. 

As noted by \citetalias{Mason2018a} and \citet{Mason2018b} the \lya\ EW distribution is likely a function of at least UV magnitude as well as the neutral fraction \citep[see][for a thorough analysis of \lya\ EW dependencies on galaxy properties]{Oyarzun2017}, so it can be difficult to compare \lya\ fraction from samples with different $\MUV$. Hence, when converting from the neutral fraction measurement in this work and \citetalias{Mason2018a} we use the model \lya\ EW distribution for $\MUV = -20$ galaxies to compare more easily with the literature values for which that is the typical median UV magnitude. For $\MUV = -20$ our \lya\ fraction limits are $f_{\textrm{Ly}\alpha} < 0.11$ (68\%), $<0.27$ (95\%). Using our sample median magnitude, $\MUV = -18.2$, the limits are not significantly different: $f_{\textrm{Ly}\alpha} < 0.08$ (68\%), $<0.24$ (95\%). Our measurements are consistent with the literature values.

We note that our inference assumes no evolution in the emitted \lya\ EW distribution at fixed UV magnitude from $z\sim6-8$, i.e. the only evolution in the \textit{observed} EW distribution is due to reionization. Whilst there may be evolution in the amount of \lya\ escaping the ISM of galaxies with increasing redshift, it is probably increasing as dust masses and HI covering fractions may decrease at higher redshifts and facilitate \lya\ escape at fixed galaxy mass \citep{Hayes2011,Oyarzun2016}. In this case we expect our model to underestimate the observed EW distribution, which would suggest an even higher neutral hydrogen fraction given our non-detections. Our model also assumes no significant evolution in the dust spatial distribution and/or CGM opacity between $z\sim6-8$, which could both reduce the \lya\ EW before the photons reach the IGM. If these effects do significantly decrease \lya\ EW between $z\sim6-8$, this could lower our constraint on the neutral fraction. In modelling the emitted \lya\ EW distribution we assume a Gaussian plus Dirac Delta function parameterisation, which has been shown to describe the \lya\ EW distribution well \citep{Oyarzun2017}. However, choosing another functional form for the distribution will not significantly change the results \citep{Treu2012,Schenker2014}.

More accurate models of \lya\ emerging from the $z>6$ ISM are required to improve our inferences. Whilst it is increasingly difficult to directly observe all of the emitted \lya\ from $z>6$ galaxies, because of the intervening neutral gas, other emission lines could be used as a diagnostic of emerging \lya. For example, \citet{Henry2018} showed that Mg\textsc{ii} emission line profiles and escape fractions closely trace those of \lya\ in Green Peas, low-redshift analogues of high redshift galaxies \citep{Jaskot2014,Yang2016}. As the IGM optical depth to Mg\textsc{ii} is much lower than for \lya, observations of Mg\textsc{ii} at $z>6$ (which will be possible with JWST) could be used infer the nature of \lya\ emission at these redshifts. 

Additionally, better knowledge of \lya\ line profiles at $z\simgt5$ are necessary to provide more informative priors on the observed FWHM for our inferences. In particular, high resolution spectroscopy ($R>4000$) is needed to resolve the narrow lines expected for UV faint galaxies \citep{Verhamme2015}, and could provide additional constraints via the evolving \lya\ profile \citep{Pentericci2018} and the prevalence of double-peaked \lya\ in the late stages of reionization \citep{Matthee2018}. 

We also assume the fraction of low redshift contaminants in our photometric sample is the same as our reference $z\sim6$ sample from \citet{DeBarros2017}. Whilst the selection techniques for the two samples are different (ours is based on photometric redshifts, \citet{DeBarros2017} uses a colour selection) our targets have extensive multi-wavelength photometry which help rule out low redshift contaminants \citep[e.g.,][]{Vulcani2017,Livermore2018}. Additionally, we use the full photometric redshift distribution from EAzY in our inference which will weight the most convincing high redshift candidates most strongly in our inference, and robustly account for contamination. With the final GLASS \lya\ candidate sample it will be possible to use the same selections for both the $z\sim6$ reference EW distribution and the $z>6$ samples for reionization inferences (K. B. Schmidt et al., in prep). 

As our inference weights sources by their photometric redshift distribution, the tightest constraints on $\xHI$ will be obtained from samples with robust redshift estimates or, ideally, spectroscopic redshifts obtained from other emission lines, and deep \lya\ EW limits. We note that the objects which constribute the most to our posterior are the objects with the highest probability ($>60\%$) of having photometric redshift at $z\sim8$ due to their SEDs, and we expect these to have consistent high redshift solutions even if the photometric redshift fitting priors are changed. The prospects for large spectroscopic samples at these redshifts is increasing: ALMA is enabling spectroscopic confirmation of $z\simgt7$ galaxies in the sub-mm \citep[e.g.,][]{Bradac2017,Laporte2017,Smit2018,Hashimoto2018,Tamura2018}, and other UV emission lines have also been confirmed \citep{Stark2015,Schmidt2016,Stark2017,Mainali2017,Mainali2018}. Future observations with JWST slitless and slit spectroscopy will be able to build large and deep spectroscopic samples of $z\simgt7$ galaxies, ideal for this type of analysis.

Understanding the differing evolution of \lya\ emission as a function of galaxy properties and environment will be key to understanding how reionization progresses. Here we have shown that a sample of intrinsically UV faint systems at $z\sim8$ (more likely to live in low density environments) show no significant \lya\ emission, and favour a mostly neutral IGM. However, \lya\ has been observed in a handful of UV bright galaxies at $z\simgt7.5$ \citep{Zitrin2015a,Oesch2015,Roberts-Borsani2016,Stark2017}. \citet{Mason2018b} showed that the observed \lya\ fraction for UV bright galaxies at $z\sim8$ could not be reproduced with standard reionization models \citep[using the EoS simulations,][]{Mesinger2016a}, even when placing them in overdense regions (which reionize early) and giving them high \lya\ velocity offsets to facilitate \lya\ IGM transmission. \citet{Mason2018b} proposed those objects have detectable \lya\ because they have unusually high emitted \lya\ EW \citep[they were certainly selected to have high nebular line EW,][]{Roberts-Borsani2016}.

Fluctuations in the UV background during reionization, for example, due to the inhomogeneous distribution of ionizing sources, could also contribute to the differing evolution of \lya\ emission from UV bright and UV faint galaxies by boosting the IGM opacity (transparency) in underdense (overdense) regions \citep{Davies2016,Becker2018}. One important missing piece in our inference is the halo environment of the LBGs. This work assumes a simple mapping between UV luminosity and halo mass. This works well in an average sense \citep{Mason2015}, but deep imaging with JWST could measure the clustering strength and scatter of galaxies in the reionization epoch \citep{Ren2018}, and be used to inform more realistic IGM simulations.

\subsection{M1149\_JD1 -- \lya\ emission at $z=9.11$?}
\label{sec:dis_JD1}

One target in our observations \citep[known as M1149\_JD1,][]{Zheng2012,Hoag2018} was recently spectroscopically confirmed at $z=9.11$ via [\OIII]$88\,\mu$m emission with ALMA observations \citep{Hashimoto2018}. Our EAzY photometric redshift distribution for this galaxy put it outside of our inference redshift range (all of the $p(z)$ is at $z>9$, see Figure~\ref{fig:pz}), so it was not used in our reionization inference. However, \citet{Hashimoto2018} also report a tentative 4$\sigma$ detection of \lya\ emission from this galaxy in X-shooter observations at 12271.5\,\AA\ with total line flux $(4.3 \pm 1.1) \times 10^{-18}$\,\fdens. \citet{Hoag2018} also targeted this galaxy with low resolution HST grism spectroscopy, including GLASS data, which covered the Lya wavelength at $z=9.11$. While they did not claim a detection, their spectra show a $\sim2.5\sigma$ feature at approximately the same wavelength and flux as \citet{Hashimoto2018}. We examined our KMOS cube and find no evidence of a feature at this wavelength. Our median $1\sigma$ flux limit for $z>9$ \lya\ in the cube is $>1.1\times10^{-18}$\,\fdens, and $>0.8\times10^{-18}$\,\fdens\ at 12271.5\,\AA.

As noted by \citet{Hashimoto2018}, if their candidate line is \lya, it is blueshifted by $\sim450$ km/s with respect to the [\OIII] emission. \citet{Hashimoto2018} suggest that \lya\ photons scattered off inflowing gas, causing it to emerge blueshifted from the galaxy's systemic velocity. Whilst blueshifts due to inflows are expected and observed for \lya\ \citep[e.g.,][]{Verhamme2006,Dijkstra2006b,Trainor2015}, at $z>6$ the IGM is opaque to emission $<1216\,$\AA, thus no \lya\ emitted bluer than its source galaxy's systemic redshift should be transmitted through the IGM \citep{Dayal2011,Dijkstra2011}. 

Observing blueshifted \lya\ requires the galaxy to sit in a large ionized bubble \citep[$\simgt500$ km/s or $\simgt400$\,kpc in radius,][]{Haiman2002}. Alternatively, the \lya\ emission could arise in a different component or merging companion of the [\OIII] emitting galaxy, similar to a $z=7.1$ galaxy observed by \citet{Carniani2017a}. The tentative emission we observe in our KMOS cube does appear spatially offset from the predicted position of the UV continuum and [\OIII] by $\sim0\farcs4$, which could provide evidence for the multi-component/merger scenario. This may also account for our slightly lower flux measurement as the emission extends to the edge of the IFU. However, the weakness of the detection and some general astrometric uncertainty in KMOS make a thorough analysis difficult. Deeper near-IR IFU observations of this galaxy would be extremely interesting to confirm and determine the nature of the \lya\ emission, and will be possible in the future with JWST NIRSpec.

We calculate the probability of observing \lya\ emission from such an object in a mostly neutral IGM using the framework of \citetalias{Mason2018a}, which modelled $p(W \,|\, \xHI, \MUV)$. Using $m_\textrm{F160W} = 25.7$ \citep{Zheng2012} we obtain $\MUV = 19.2 - 2.5\log_{10}(10/\mu)$, $EW = 4\pm2$\,\AA\ for our measured flux and $EW = 11\pm3$\,\AA\ from the measurement by \citet{Hashimoto2018}. Using these measurements we calculate $p(W = 4\pm2 \,\textrm{\AA} \,|\, \xHI > 0.76, \MUV = -19.2) < 0.05$, while $p(W = 11\pm3 \,\textrm{\AA} \,|\, \xHI > 0.76, \MUV = -19.2) < 0.03$. In fact, the total probability of observing \lya\ from this galaxy with EW $>4\pm2$\,\AA\ if $\xHI > 0.76$ is $\simlt0.5$: low \lya\ EW are expected and consistent with a mostly neutral IGM.

We note that our calculations assume the \lya\ is emitted close to systemic velocity (i.e., assuming that the \lya\ comes from another component). Obviously if the galaxy does sit in an ionized bubble the probability of seeing emission would be higher. But we note that assuming emission is emitted at systemic velocity the probability of detecting the emission is not negligible, and thus this detection is still consistent with a mostly neutral IGM at $z>8$.

\subsection{Other UV emission lines at $z\sim8$}
\label{sec:dis_otherUV}

With \lya\ increasingly suppressed at $z>6$, rest-frame UV emission lines can be used to spectroscopically confirm high-redshift LBGs. These lines can also be used as diagnostics for the stellar populations and physical conditions present in these high-redshift galaxies. Our KMOS observations cover the wavelength range where NV$\lambda1238,1242$ and \CIV$\lambda1558,1551$ can be observed, and we briefly discuss our upper limits on the EW of these lines.

NV$\lambda1238,1242$ can arise due to stellar winds, particularly from very young stars \citep{Shapley2003,Jones2012}, or from H\textsc{ii} regions if powered by an AGN or radiative shocks. Of the three $z>7$ galaxies detected to-date with tentative NV emission ($S/N \sim 4$) all have been UV bright galaxies, where AGN activity could plausibly be powering NV emission \citep{Tilvi2016,Laporte2017a,Mainali2018}. In our KLASS $7.2 < z_\mathrm{phot} < 8.8$ sub-sample (Section~\ref{sec:reionization_phot}), the median NV EW upper limit is $<60$\,\AA. As our sample comprises intrinsically faint galaxies, which are less likely to have strong AGN activity, it is not surprising we do not detect strong NV emission.

Nebular \CIV$\lambda1558,1551$ emission has been observed in two galaxies at $z>6$ \citep{Stark2015,Schmidt2017a,Mainali2017}. We observed \CIV\ in the $z=6.11$ galaxy with KMOS and describe our observations in more detail in Appendix~\ref{app:CIV}. The \CIV\ emission can be powered by either AGN activity or extremely metal poor stars. Limits on other UV lines in these objects find low metallicity stars are a more likely source of the hard photons needed to produce \CIV\ emission, rather than AGN. The two galaxies are also both UV faint galaxies ($\MUV \simlt -20.2$) and \citet{Mainali2018} has suggested that there is anti-correlation between UV luminosity and \CIV\ EW, which could arise if the lowest luminosity (mass) systems are more metal-poor. 

Our KLASS observation provide a large additional sample of UV faint galaxies which can place new limits on \CIV\ emission. In our KLASS $7.2 < z_\mathrm{phot} < 8.8$ sub-sample (Section~\ref{sec:reionization_phot}), the median \CIV\ EW upper limit is $<74$\,\AA. In the three most UV faint systems with $P(7.2 < z_\mathrm{phot} < 8.8) > 0.6$, M0416\_22746, RXJ1347\_656, M0416\_1997 (all with $\MUV \sim -17.5$), the \CIV\ upper limits are $<62$\,\AA, $<22$\,\AA, and $<51$\,\AA\ respectively. These upper limits are comparable to, and in one case below, the \CIV\ detection presented by \citet{Stark2015} in a $\MUV \sim -19$ galaxy (with $EW_\textsc{civ} \approx 38$\,\AA), and so suggest that the proposed anti-correlation between UV luminosity and \CIV\ EW may not be so simple.

\subsection{Background limited observations with KMOS}
\label{sec:dis_KMOS}

\begin{figure*} 
\centering
\includegraphics[width=0.99\textwidth, clip=true]{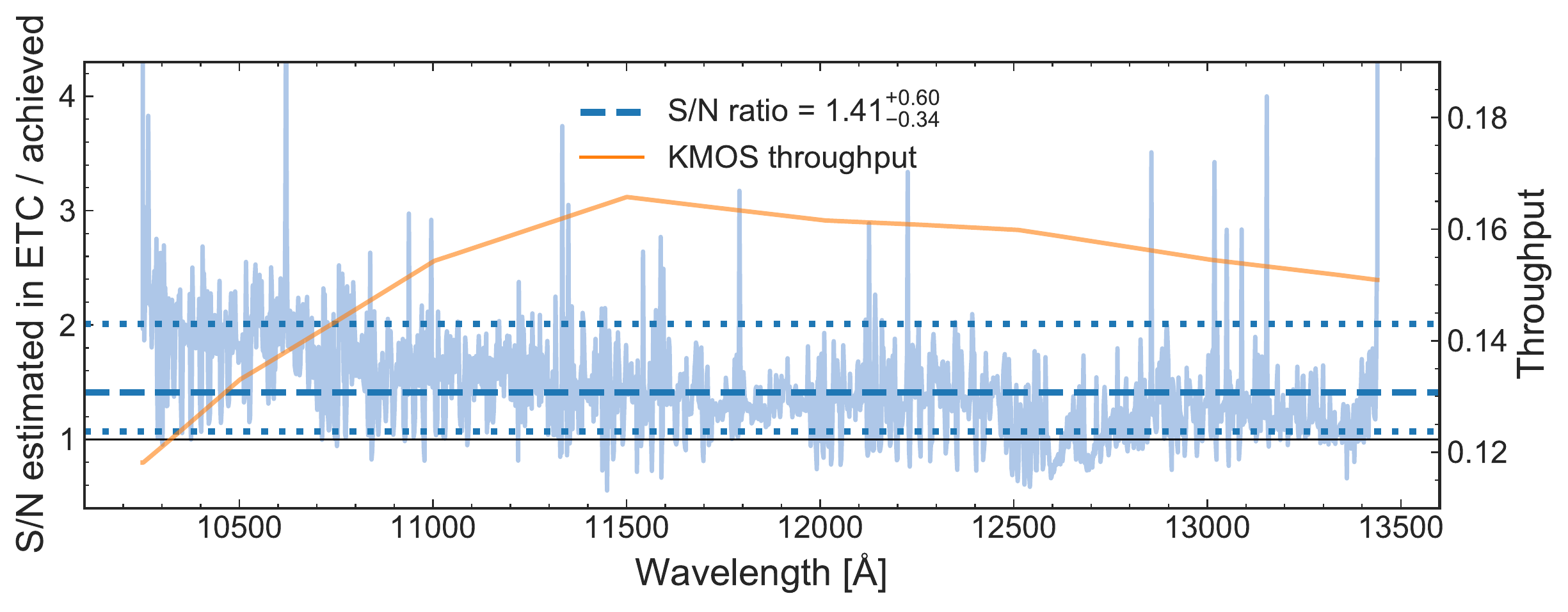}
\caption{Comparison of our deepest exposure, 11 hours in in RXJ1347, with 450 second DITs, with the ESO KMOS ETC using the same exposure times. We compare the $5\sigma$ flux limits from our data and the ETC as a function of wavelength, assuming emission lines are spatially and spectrally unresolved. We divide the ETC estimated noise by $\sqrt{2}$ to account for the noise introduced by `A-B' sky subtraction. The pale blue solid line shows the ratio of the ETC estimated S/N to our achieved S/N. The blue dashed (dotted) horizontal lines show the median ($16-84\%$ range) of the S/N ratio over the whole YJ range. The orange line shows the KMOS throughput for comparison.}
\label{fig:ETC}
\end{figure*}

Optical and near-IR IFU observations have provided revolutionary 3D information about the structure and kinematics of galaxies out to $z\sim2$ \citep{ForsterSchreiber2009,Epinat2009,Wisnioski2015,Stott2016,Genzel2017} and revealed diffuse \lya\ halos around $z\simlt6$ galaxies \citep{Bacon2014,Karman2016,Wisotzki2016,Leclercq2017}.

In KLASS we have provided the first deep NIR IFU observations of $z\simgt7$ galaxy candidates. Whilst we did not make any $5\sigma$ detections of \lya\ it is important to understand how this depended not only on the selection of our targets and the opacity of the IGM to \lya\ at $z\sim8$, but on the sensitivity of KMOS. In our long integrations we have pushed KMOS to the limits of its sensitivity to search for faint emission lines in near-IR IFU cubes, in wavelength regions dominated by OH sky emission lines. Using our deep observations we provide an assessment of the performance of KMOS for background-limited observations.

As described in Section~\ref{sec:obs_reduction} we performed additional sky subtraction routines after running the ESO pipeline to reduce residuals around bright OH lines. We also found the pipeline underestimated the noise in cubes by a factor $\sim1.2$ and performed additional rescaling of the noise as a function of wavelength using the RMS noise obtained from the flux cubes.

One key question is how well the instrument performs relative to the predictions based on its instrumental capabilities. We can compare S/N estimated by the KMOS ETC\footnote{\url{https://www.eso.org/observing/etc/bin/gen/form?INS.NAME=KMOS+INS.MODE=lwspectr}} to our achieved S/N to assess its performance. We take the $5\sigma$ flux density limits as a function of wavelength for our deepest exposure, 11 hours in RXJ1347, (shown in Figure~\ref{fig:fluxlim}) and calculate the S/N as estimated by the ETC. We use our flux calibration based on observations of standard stars to convert flux to e$^-$/s and rescaled by the wavelength-dependent sky transmission and KMOS throughput curve (both obtained through the KMOS ETC webpage). We use the following ETC settings which are comparable to those of our observations: line FWHM$_\textrm{spec}=4$\,\AA\ (unresolved); point source; seeing $0\farcs6$; airmass: 1.50, Moon illumination FLI: 0.50, Moon-target separation: 45 degrees, PWV: $<2.5$\,mm. We calculate the S/N in an aperture with radius equal to the seeing FWHM $\sim 0\farcs6$.

At every wavelength, the estimated S/N is:
\BE \label{eqn:ston}
\frac{S}{N} = \frac{\sqrt{NDIT} \times S_\textrm{source}}{\sqrt{S_\textrm{source} + S_\textrm{bkg} + n_\textrm{spat}(\textrm{DC}\times \textrm{DIT} + \textrm{RON}^2)}}
\EE
where for RXJ1347 NDIT $=88$ is the number of DITs, of length DIT $=450$ seconds. The KMOS dark current (DC) is $0.01$ e$^{-}$/pixel/s and the read-out noise is $3.5$ e$^{-}$/pixel/DIT. The aperture corresponds to $n_\textrm{spat} = 25$ spatial pixels and the calculation is done at the peak wavelength pixel. We use the online ETC to generate the background flux $S_\textrm{bkg}$ in e$^{-}$/DIT as a function of wavelength, convolved with the instrumental resolution, given our input settings described above. We then calculate the estimated S/N using Equation~\ref{eqn:ston} at every wavelength using our 5$\sigma$ flux density limits as the source flux.

In Figure~\ref{fig:ETC} we show a comparison of the ETC estimated S/N as a function of wavelength for the line fluxes corresponding to our $5\sigma$ limits. We plot the S/N estimated by the pipeline divided by 5 to show how the achieved S/N compares to the predicted S/N from the ETC. The public ETC does not account for noise due to sky subtraction routines. Assuming all DITs have equal noise $\sigma$, for `A-B' frames the noise should be $\sqrt{2}\sigma$. Thus in Figure~\ref{fig:ETC} we also divide the ETC estimate by a factor $\sqrt{2}$ for a fairer comparison with our data. We find that the ETC S/N is a median $\sim1.4\times$ higher than our achieved values, and this overestimate is highest for wavelengths $\simlt 11500$\,\AA, where the ETC estimate can be $\sim1.6-1.8\times$ higher. 

As shown in Figure~\ref{fig:ETC}, the KMOS YJ throughput is known to decrease at $\simlt 11500$\,\AA\ but our results suggests that the YJ grating is less sensitive in the blue for background-limited observations than expected.

Unfortunately this corresponds to \lya\ redshifts $z\simlt8.5$, where we expect to find the majority of our targets. Using the S/N estimated from the ETC in planning our observations likely led us to overestimate the line sensitivity of KMOS for our targets. Most of the GLASS \lya\ candidates we assigned to KMOS IFUs had tentative detections in the \HST\ grisms. Thus a key aim of the deeper KMOS observations was to confirm these emission lines. While our deepest $1\sigma$ flux limit in our KMOS sample is $0.8\times10^{-18}$\,\fdens, deeper than the $1\sigma$ flux limit in GLASS ($5\times10^{-18}$\,\fdens), we did not detect any emission from the tentative GLASS \lya\ candidates with KMOS, suggesting that some of the HST grism lines were spurious noise fluctuations. A thorough comparison of the GLASS and KLASS observations, in combination with other follow-up at Keck, to determine the HST grism purity and completeness will be discussed in a future paper.

We advise any future KMOS users planning observations of faint targets to take into consideration both the additional noise from sky subtraction when using the KMOS ETC, and the lower than expected performance at the blue end of YJ. However, we find better agreement with the ETC estimates at redder wavelengths, demonstrating that KMOS YJ is performing well at $\simgt 11500$\,\AA.

\section{Summary and Conclusions}
\label{sec:conc}

We have presented an analysis of reionization epoch targets from KLASS, a large ground-based ESO VLT/KMOS program following up sources studied in the HST grism survey GLASS. Our main conclusions are as follows:

\begin{enumerate}
\item The median $5\sigma$ flux limit of our survey is $4.5 \times 10^{-18}$\,\fdens. We determine our spectroscopic survey to be 80\% complete over the full wavelength range for $7.2 \simlt z \simlt 10.1$ spatially unresolved \lya\ emission lines with flux $\simgt 5.7\times10^{-18}$\,\fdens, centred within $0\farcs8$ of the IFU centre and with intrinsic line FWHM$_\textrm{spec}\simlt 250$\,\kms. Our observations are more complete to \lya\ emission that may be spatially offset and/or extended compared to the UV continuum than typical slit-spectroscopy surveys.
\item Of the 52 $z\simgt7$ candidate targets observed, none have confirmed \lya\ emission, including those with candidate lines detected in the HST grisms. No other UV emission lines are detected at $z>7$. We detect \CIV\ emission in one image of a previously known \CIV\ emitter at $z=6.11$.
\item We define a sub-sample of 29 targets with a homogeneous photometric selection of $7.2 < z_\textrm{phot} < 8.8$ for a Bayesian inference of the IGM neutral hydrogen fraction. The median \lya\ flux limit for our sample is $3.6\times10^{-18}$\,\fdens\ and the median \lya\ EW upper limit is $58$\,\AA. Combining our sub-sample with 8 previously observed $z\sim8$ LBGs from the BoRG survey \citep{Trenti2011,Treu2013,Schmidt2014a} we obtain a lower limit on the IGM neutral hydrogen fraction at $z=7.9\pm0.6$, $\xHI > 0.76$ (68\%) and $\xHI > 0.46$ (95\%).
\item Our constraint favours a late reionization consistent with models where ultra-faint galaxies contribute significantly to reionization, with an ionizing photon escape fraction $\langle f_\textrm{esc} \rangle \simlt 15\%$.
\end{enumerate}

Our KMOS observations provide more evidence of a predominantly neutral IGM at $z\sim8$. To make more precise constraints on the timeline of reionization will require larger samples of LBGs with precise photometric (or even better, spectroscopic) redshift estimates, more informative priors on \lya\ FWHM, and deep spectroscopic limits on \lya. Forthcoming deep spectroscopic observations with JWST \citep[e.g.,][]{Treu2017} will provide ideal samples for future inferences on reionization.

\section*{Acknowledgements}

The authors thank the referee for their constructive comments. We thank Trevor Mendel and Owen Turner for useful discussions related to KMOS reductions for faint sources, and T. Mendel for sharing the readout correction code. We thank Andrei Mesinger for providing \lya\ optical depths from the EoS simulations.

C.M. acknowledges support by NASA Headquarters through the NASA Earth and Space Science Fellowship Program Grant NNX16AO85H, and through the NASA Hubble Fellowship grant HST-HF2-51413.001-A awarded by the Space Telescope Science Institute, which is operated by the Association of Universities for Research in Astronomy, Inc., for NASA, under contract NAS5-26555. This work was supported by the HST GLASS grant GO-13459, the HST BoRG grants GO-12572, 12905, 13767 and 15212, and HST-AR-13235 and HST-AR-14280.

This work was based on observations collected at the European Organisation for Astronomical Research in the Southern Hemisphere under ESO program 196.A-0778; and on observations made with the NASA/ESA Hubble Space Telescope, obtained at STScI. We are very grateful to all ESO and STScI staff who have assisted in these observations. 

This work utilises gravitational lensing models produced by PIs Brada{\v c}, Natarajan \& Kneib (CATS), Merten \& Zitrin, Sharon, Williams, Keeton, Bernstein and Diego, and the GLAFIC group. This lens modelling was partially funded by the HST Frontier Fields program conducted by STScI. The lens models were obtained from the Mikulski Archive for Space Telescopes (MAST).

\textit{Software}: IPython \citep{Perez2007}, matplotlib \citep{Hunter2007}, NumPy \citep{VanderWalt2011}, SciPy \citep{Oliphant2007}, Astropy \citep{Robitaille2013}, QFitsView (\url{http://www.mpe.mpg.de/~ott/QFitsView/}).

\bibliography{library}
\bibliographystyle{mnras}
\appendix

\section{Independent confirmation of CIV emission at $z=6.11$}
\label{app:CIV}

\begin{figure*} 
\centering
\includegraphics[width=0.99\textwidth, clip=true]{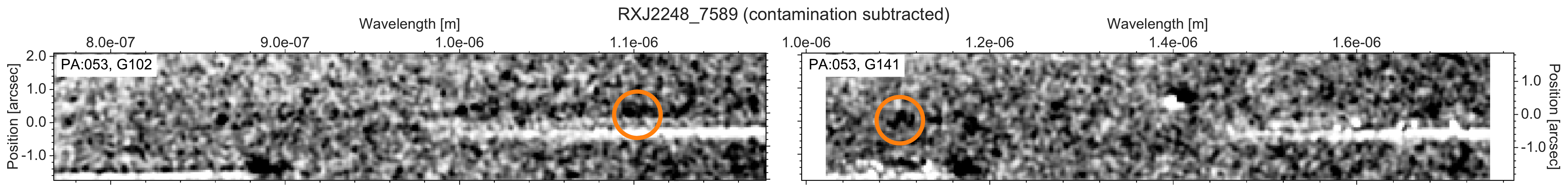}
\includegraphics[width=0.99\textwidth, clip=true]{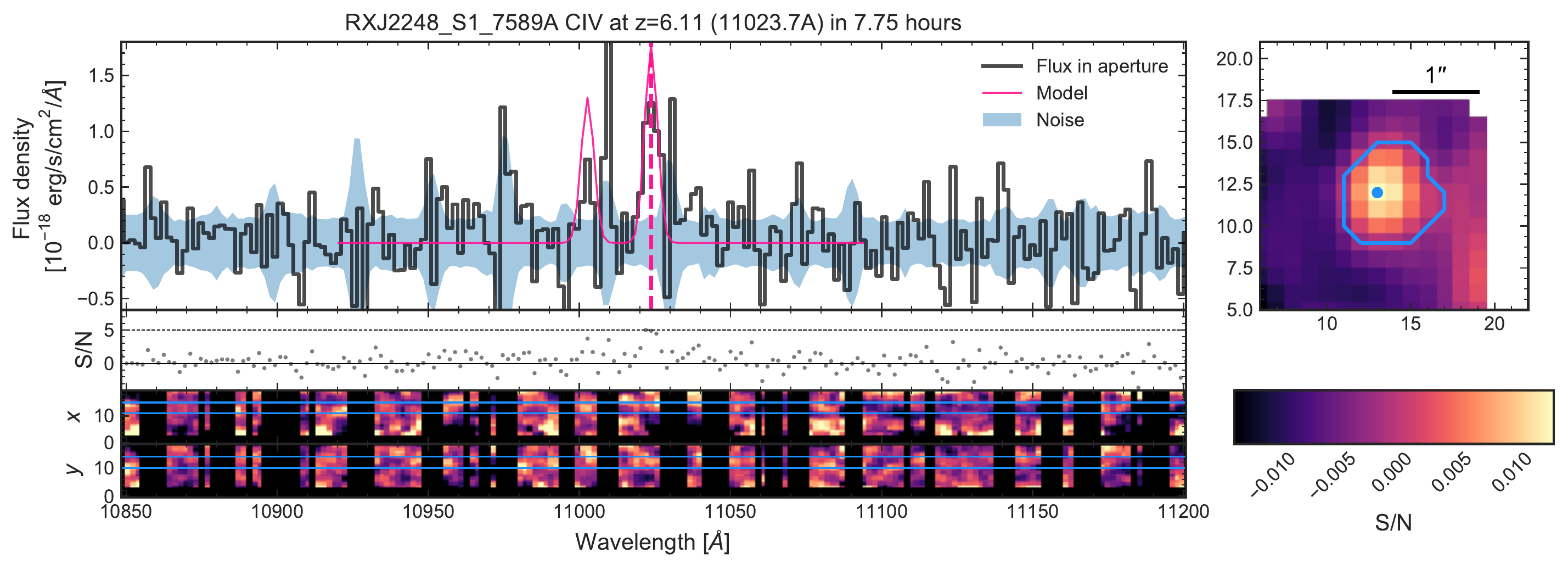}
\caption{\textbf{Upper panels:} GLASS \HST\ grism spectra taken at 2 position angles (PA) in G102 and G141. Positive flux is shown in black, white is negative flux. The candidate emission line is within the orange circle. \textbf{Lower left panel}: KLASS KMOS spectra of the same object, centred at the wavelength of the GLASS candidate emission. The top left panel shows a 1D spectrum (flux -- black line, noise -- blue shaded region) extracted in an aperture shown by the blue lines in the 2D postage stamp image on the right (an aperture containing the brightest 10\% of pixels, with area 0.8 sq arcsec). The wavelength of the 1551\,\AA\ emission line is shown with the pink dashed vertical line. The model for the doublet using 2 Gaussians, extracting voxels in a 1.2 sq arcsec aperture (shown as the blue contour on the 2D image - lower right panel), is shown as a pink solid line. The middle panel shows the S/N at each wavelength pixel. The lower spectra show simulated slit spectra along the $x$ and $y$ directions, with the same aperture plotted as blue horizontal lines. Regions with strong sky emission are masked. The spectra are smoothed with a 3D Gaussian kernel with line spread function FWHM$_\textrm{spec}$ equal to the instrumental resolution (4\,\AA) and point spread function FWHM$_\textrm{spat}$ equal to the seeing ($0\farcs6$). \textbf{Lower right panel:} 2D flux postage stamp image of the emission summed over a 10\,\AA\ wavelength range centred at the wavelength shown by the pink dashed vertical line in the left panel. The emission is clearly spatially compact. The blue contour shows the spatial aperture used to extract the 1D spectra. The S/N colourbar refers to the 2D slit spectra. We have performed an additional residual sky subtraction to the KLASS cube by subtracting a median 1D spectrum obtained in spaxels away from the emission line.}
\label{fig:CIV}
\end{figure*}

As well as \lya\ candidates we also targeted 3 images of a multiply-imaged $z=6.11$ galaxy in RXJ2248 to observe \CIV$\lambda$1548,1551 emission. This multiple-image system has been detected with \lya\ emission by \citet{Boone2013,Balestra2013,Monna2014,Karman2015,Schmidt2016} and \citet{Mainali2017}. Detections of \CIV\ and \OIII]$\lambda$1666 emission in one of the images were also presented by \citet{Mainali2017}, and \CIV\ detections and limits in all of the images by \citet{Schmidt2017a}. The presence of these highly ionised lines and lack of observed \HeII$\lambda$1640 indicate this object has a hard ionizing spectrum, but unlikely to be dominated by an AGN.

In Figure~\ref{fig:CIV} we show our GLASS and KMOS spectra for the brightest image E. An emission line is seen in the KMOS data at the same wavelength as \citet{Mainali2017}, and appears to be spatially compact with size approximately equal to our seeing ($\sim0\farcs6$). However, due to the lower spectral resolution of our data \citep[$R\sim3400$ compared to $R\sim6000$ in][]{Mainali2017} the emission line is overlapping with an adjacent skyline, adding some uncertainty to the extracted parameters. 

We fit a model with 2 Gaussian emission lines to an extracted 1D spectrum, weighted by the inverse variance, fixing FWHM$_\textrm{spec} = 5$\,\AA\ (close to the instrumental resolution, 4\,\AA) and allowing flux and the wavelength position of the doublet centre to vary. We note that the spatial aperture we use to extract the spectrum (1.2 sq arcsec) is slightly larger than the one used to plot Figure~\ref{fig:CIV} (0.8 sq arcsec), to ensure we recover the full flux. We use the smaller aperture in the plot to maximise the plotted S/N and for comparison with the line identification technique described in Section~\ref{sec:results_lines} which also uses 0.8 sq arcsec apertures.

The strongest peak is fit at 11023.7\,\AA\ which \citet{Mainali2017} assigned to the 1551\,\AA\ emitted peak. There is a weaker peak at 11002.5\,\AA\ which would corresponds to the 1548\,\AA\ emitted peak. We note that in the 1D spectrum there does appear to be a peak redward of the strong peak, around 11040\,\AA\ which could alternatively be the 1551\,\AA\ emitted peak. However, the peak separation to the 11040\,\AA\ line is too low for this to be part of the doublet and by inspection of the 2D emission postage stamps in the KMOS cubes we see the 11002.5\,\AA\ flux is spatially coincident with the 11023.7\,\AA\ emission, whilst the 11040\,\AA\ flux is more spatially uniform and thus likely to be spurious noise peak. Thus our assignment of the doublet wavelengths is consistent with the observations of \citet{Mainali2017}.

Using our model we find a total line flux of $(1.6\pm0.3) \times 10^{-17}$\,\fdens. We find a flux ratio of \CIV$\lambda$1548:\CIV$\lambda$1551 = 0.7:1 which is much lower than the theoretically motivated expected value of 2:1 \citep[e.g.,][]{Flower1979}. Low mass metal-poor galaxies at $z\sim2-3$ have been observed with flux ratios both comparable to the theoretical value \citep{Vanzella2016a,Caminha2016} and closer to a 1:1 ratio \citep{Christensen2012,Stark2014}. 

However, the confusion with the adjacent skylines makes it difficult to measure an accurate flux. Our measured total flux is lower than those measured by \citet{Mainali2017} and \citet{Schmidt2017a}, but consistent with the value measured by \citet{Schmidt2017a} within $2\sigma$. We also find a tentative (S/N $\sim3$) detection at the wavelength where O\textsc{iii}]$\lambda$1666 was identified by \citet{Mainali2017} in our data (11837.1\,\AA).

\section{Testing the selection function of our final sub-sample}
\label{app:phot}

\begin{figure*} 
\centering
\includegraphics[width=0.49\textwidth, clip=true]{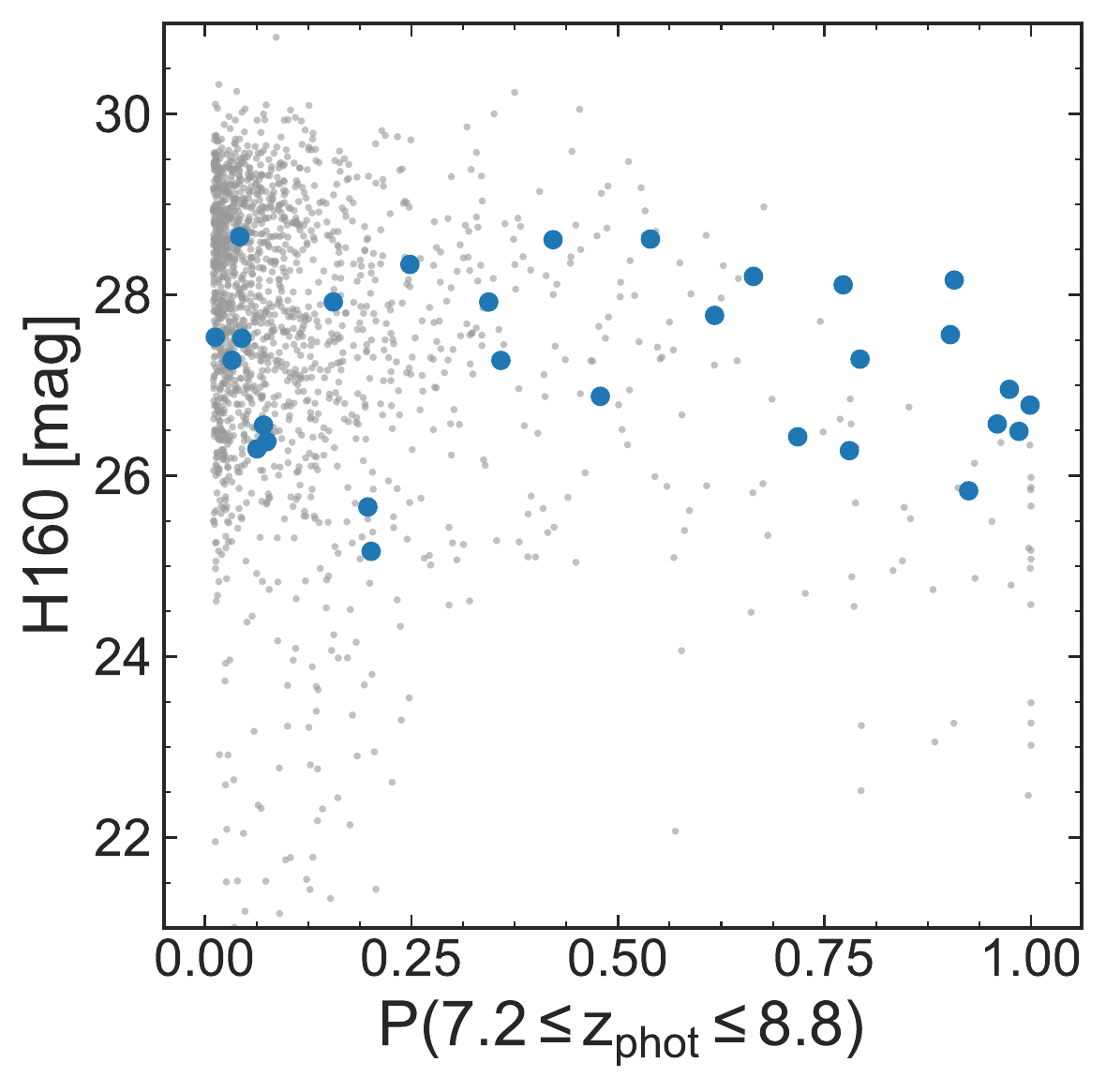}
\includegraphics[width=0.49\textwidth, clip=true]{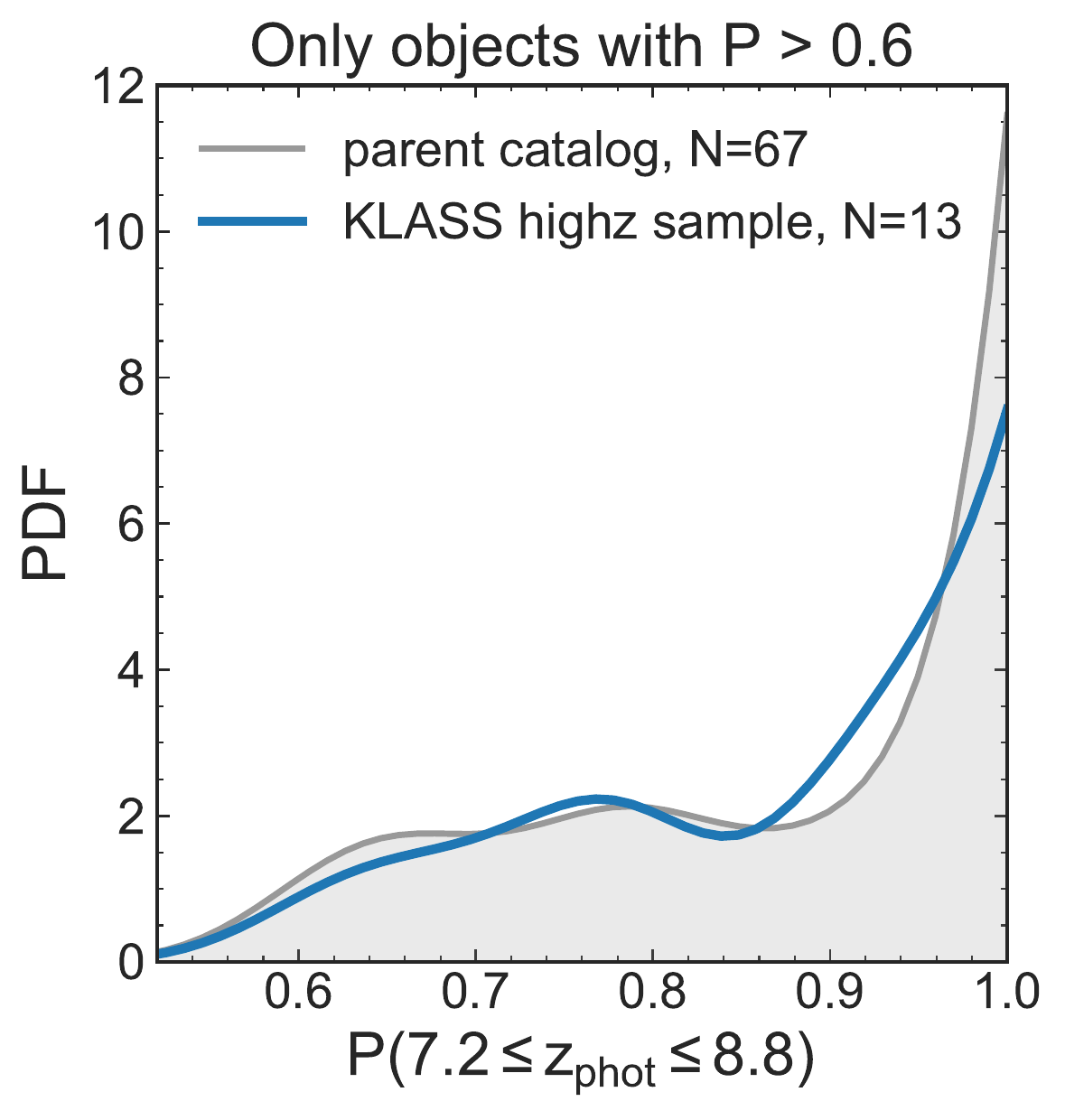}
\caption{\textbf{Left}: Total probability (from the EAzY photometric redshift distributions) of objects being within our redshift range of interest ($P(7.2\geq z_\mathrm{phot} \geq 8.8) > 0.01$) versus their F160W apparent magnitude. The KLASS sub-sample used for the inference is shown as large blue circles, objects from the full parent photometric catalogues are shown as small grey dots. Our sample is skewed towards higher probability of $7.2\geq z_\mathrm{phot} \geq 8.8$ compared to the parent sample, with a smaller range in F160W magnitude. \textbf{Right}: Probability distribution functions of $P(7.2\geq z_\mathrm{phot} \geq 8.8)$ values of all objects with $P(7.2\geq z_\mathrm{phot} \geq 8.8) > 0.6$ from our KLASS sub-sample (blue) and the parent catalogues (grey). The distributions are plotted using a Gaussian kernel density estimator \citep{Rosenblatt1956,Parzen1962}. The two distributions are very similar, demonstrating our KLASS sub-sample is drawn randomly from the parent catalogue sample, and thus is not a biased sample in terms of photometric redshift distribution, despite being constructed after the observations were taken.}
\label{fig:app_phot_pinrange}
\end{figure*}

\begin{figure*} 
\centering
\includegraphics[width=0.54\textwidth, clip=true]{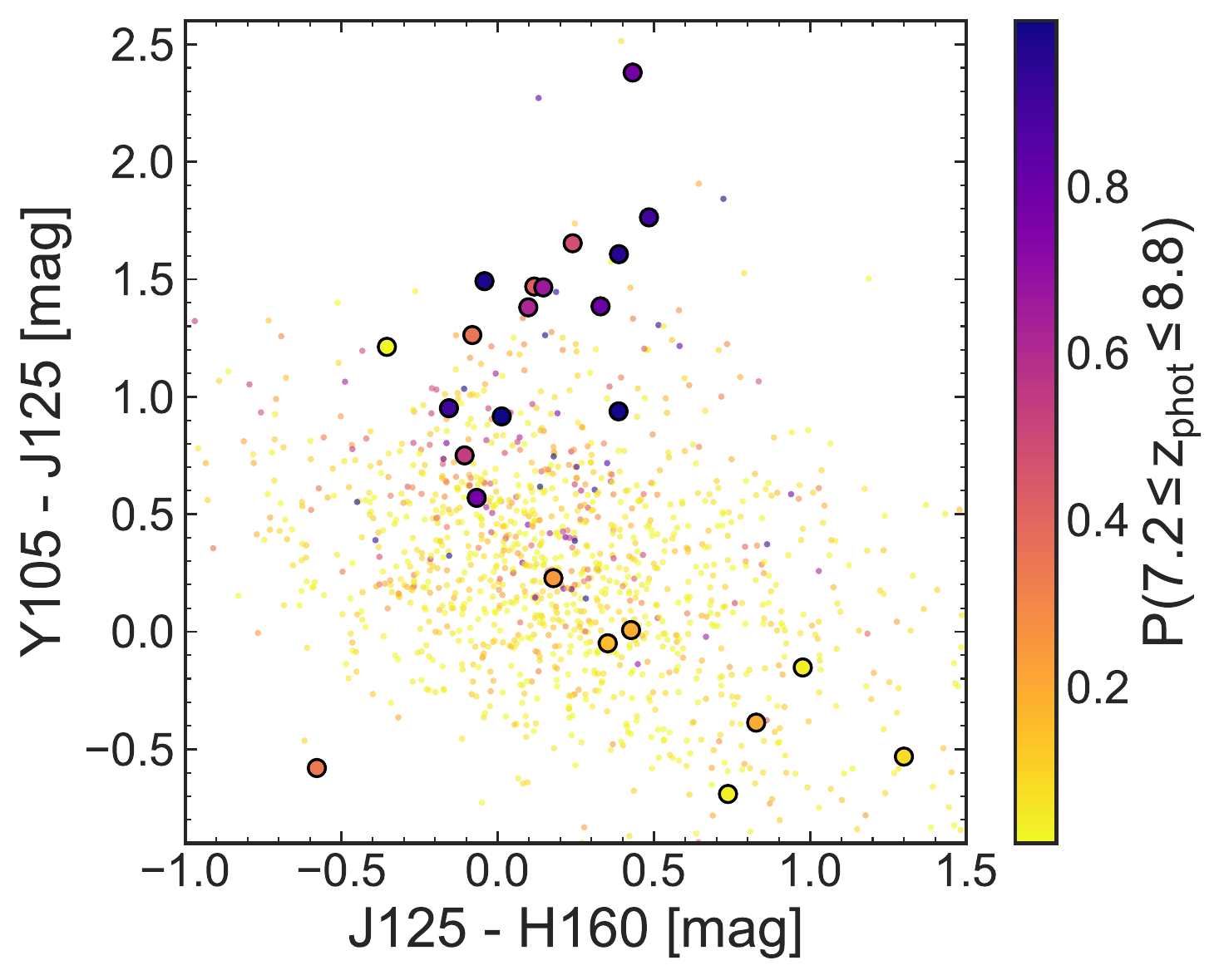}
\includegraphics[width=0.44\textwidth, clip=true]{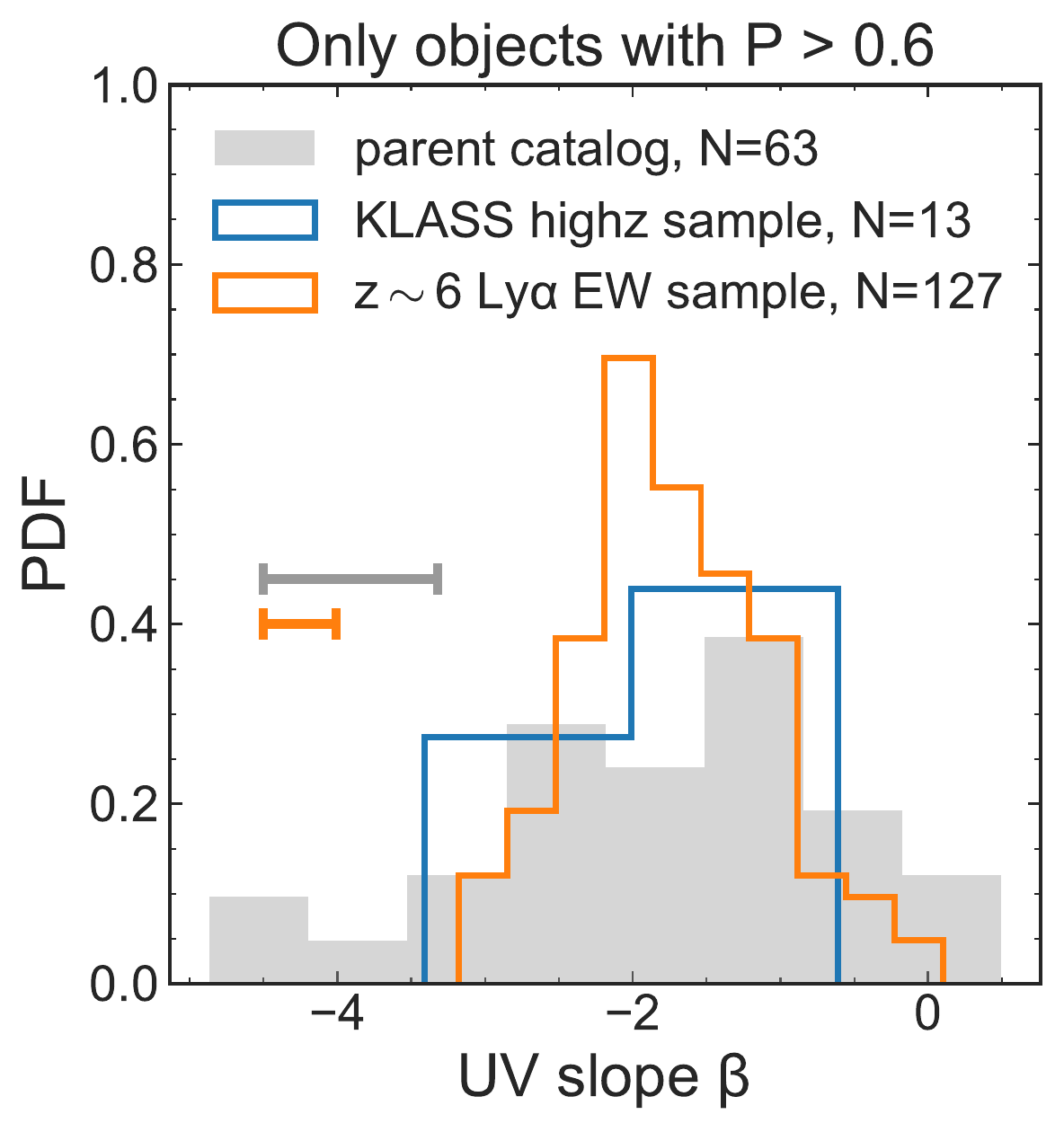}
\caption{\textbf{Left}: F125W - F160W colour versus F105W - F125W colour for our KLASS inference sub-sample (large circles with black edges) and the parent catalogues (small points). The markers are colour-coded by their $P(7.2\geq z_\mathrm{phot} \geq 8.8)$ so that the darkest points are the most likely to be within that redshift range. The most probable objects in the KLASS sub-sample appear to be in the same region of colour-colour space as the full parent catalogue. \textbf{Right}: Probability distribution functions of UV $\beta$ slope values of all objects with $P(7.2\geq z_\mathrm{phot} \geq 8.8) > 0.6$ from our KLASS sub-sample (blue) and the parent catalogues (grey). As in Figure~\ref{fig:app_phot_pinrange} the two distributions are very similar, demonstrating our KLASS sub-sample is not a strongly biased sample in terms of colour, despite being constructed after the observations were taken. We also plot the $\beta$ distribution for the $z\sim6$ \citep{DeBarros2017} sample used to create the intrinsic \lya\ EW distribution for our inference (orange). Median errorbars for the GLASS parent sample and \citet{DeBarros2017} sample are shown in grey and orange respectively. Using a Kolmogorov–Smirnov test we find the \citet{DeBarros2017} sample, KLASS high-$z$ sample, and GLASS parent catalogue $\beta$ distributions are all consistent with being drawn from the same population.}
\label{fig:app_phot_color}
\end{figure*}

As described in Section~\ref{sec:reionization_phot} we define a selection function based on the photometric redshift distributions to select a sub-sample of KLASS targets to use in the reionization inference. In this appendix we demonstrate the the sub-sample was not a biased selection from the final parent catalogues.

Objects in the sub-sample must have a match in our final deep photometric catalogues: for A2744, M0416 and M1149 we used the ASTRODEEP catalogues \citep{Castellano2016b,Merlin2016,DiCriscienzo2017}. For M2129, RXJ1347 and RXJ2248 we created our own catalogues based on the ASTRODEEP methodology. And the objects must have $P(7.2\geq z_\mathrm{phot} \geq 8.8) > 0.01$ based on photometric redshift probability distributions calculated using EAzY \citep{Brammer2008}. We use the full photometric redshift distribution in our inference to robustly weight objects based on their probability of being in our redshift range of interest (Section~\ref{sec:inference} and Appendix~\ref{app:inference}).

However, given that we construct this sub-sample \textit{after} the observations were taken, we must check that the objects we observed were not a biased sample from the final catalogues. There are many more objects in the final catalogues which were not observed in KMOS so it could be that the KLASS targets are a biased sample of the final catalogue.

In Figure~\ref{fig:app_phot_pinrange} we show the distribution of $P(7.2\geq z_\mathrm{phot} \geq 8.8)$ values of our KLASS sub-sample and the parent catalogues. The parent catalogues have many more objects with low $P(7.2\geq z_\mathrm{phot} \geq 8.8)$, showing that our sub-sample is skewed towards objects which are most likely to be within that redshift range, demonstrating that our initial $z_\mathrm{phot}>7.2$ target selection was good. In the right panel we show the probability distribution of objects in the parent catalogues and our sub-sample, using only objects with $P(7.2\geq z_\mathrm{phot} \geq 8.8) > 0.6$. As noted previously, we do not expect all the LBGs to have $P(7.2\geq z_\mathrm{phot} \geq 8.8) = 1$ as there are often degeneracies in the photometry that make $z_\mathrm{phot}\sim1-2$ solutions possible. Here we see that the distributions are consistent and demonstrate that the best targets in KLASS sub-sample (which weight most in the reionization inference) are drawn randomly from the best targets in the parent catalogues.

We also want to be sure our sub-sample is drawn randomly from the parent sample in terms of photometry. In particular, if we had selected only the reddest objects for our KMOS targets, our sub-sample could be biased towards dustier galaxies. As dust can significantly attenuate \lya\ emission \citep{Hayes2011} this would mean our non-detections of \lya\ could be due to stronger dust attenuation as well as reionization. However, we demonstrate in Figure~\ref{fig:app_phot_color} that the $\mathrm{F125W} - \mathrm{F160W}$ colour distribution and the UV slope ($\beta$, where flux $f_\lambda \propto \lambda^\beta$) distribution of our sub-sample is consistent with that of the parent catalogues. We measure UV slopes for the GLASS and KLASS objects by fitting to the $\mathrm{F125W}$ and $\mathrm{F160W}$ magnitudes \citep{Castellano2012}. The parent catalogues contain a small fraction ($<10\%$) of redder objects that are not present in our KLASS sub-sample, but given the small number of objects in our sub-sample missing this small fraction of redder objects is expected. 

In addition, in the right panel of Figure~\ref{fig:app_phot_color} we compare the UV slope distribution of the $z\sim6$ EW calibration sources used for the inference EW models \citep{DeBarros2017} to our GLASS and KLASS catalogues. Again, we want to be sure that the \citep{DeBarros2017} sample and KLASS sample are similar as we assume that these $z\sim6$ sources are a good proxy for the $z\sim8$ galaxies in our inference. We plot the UV slope distributions and the median errors on the $\beta$ measurements. For the KLASS $z\sim8$ objects we can only fit the slopes using $\mathrm{F125W}$ and $\mathrm{F160W}$ magnitudes, so these measurements have large uncertainties. Using a Kolmogorov–Smirnov test we find the \citet{DeBarros2017} sample, KLASS high-$z$ sample, and GLASS parent catalogue $\beta$ distributions are all consistent with being drawn from the same population. Therefore, differences in the \lya\ transmission from $z\sim6$ to $z\sim8$ due to dust absorption are likely to be negligible. Spectral coverage at $>2\,\mu$m with JWST will hugely improve constraints on UV slopes for $z\simgt6$ galaxies and enable a better understanding of how dust mediates \lya\ escape at high redshifts.

\section{Reionization Inference}
\label{app:inference}

This appendix derives the posterior distribution for the IGM neutral fraction, $\xHI$, extending the framework of \citetalias{Mason2018a} from single EW measurements to an input flux density spectrum as a function of wavelength, and galaxy UV apparent magnitude and gravitational lensing magnification.

\subsection{Likelihood at one spectral pixel}
\label{app:inference_likei}

We want to obtain the likelihood of observing a flux density spectrum $\{f\}$ with flux $f_i$ at wavelength pixel $i$ given our the neutral fraction $\xHI$ and the properties of that observed galaxy.

The likelihood of measuring flux density $f_i$ at wavelength $\lambda_i$ given that the photons originate at $z_d$ from a galaxy with apparent magnitude $m$ and travel through an IGM with neutral fraction $\xHI$ is
\BE \label{appeqn:inference_linelike}
\begin{split}
p&(f_i \,|\, \xHI, m, \mu, z_d, \mathrm{FWHM}) = \\
&\int_0^\infty dW \, p(f_i \,|\, W, m, z_d, \mathrm{FWHM}) \, p(W \,|\, \xHI, m, \mu, z_d)
\end{split}
\EE
Including Gaussian errors in the spectra (with error $\sigma_i$  at spectral pixel $i$), the probability of measuring flux density $f_i$ at spectral pixel $i$ is given by a Gaussian distribution at each spectral pixel, with mean given by the model flux density for a given equivalent width and standard deviation $\sigma_i$:
\BE \label{appeqn:inference_pEL}
\begin{split}
p&(f_i \,|\, W, m, z_d, \mathrm{FWHM}) ={} \\
& \frac{1}{\sqrt{2\pi}\sigma_i}\exp{\left[-\frac{ \left( f_i - f_\mathrm{mod}(\lambda_i,W,m, z_d, \mathrm{FWHM}) \right)^2 }{2\sigma_i^2}\right]}
\end{split}
\EE
The total flux of the model emission line is given by $F_{tot} = Wf_{cont}(1+z)$. For simplicity, we model emission lines as Gaussians, so that the flux density produced at a single spectral pixel $i$ by an emission line at pixel $d$ (at wavelength $\lambda_d = \lambda_\alpha(1+z_d)$) is:
\BE \label{appeqn:inference_linemod}
\begin{split}
f_\mathrm{mod}&(\lambda_i, W, m, z_d, \mathrm{FWHM}) ={} \\
& \frac{W f_{cont}(m,z_d)(1+z_d)}{\sqrt{2\pi}\sigma_\lambda} \exp{\left[-\frac{ \left( \lambda_i -\lambda_d \right)^2}{2\sigma_\lambda^2}\right]}
\end{split}
\EE
where $m$ is the observed apparent magnitude of the source, and $\sigma_\lambda = \mathrm{FWHM}/2.355$ is the spectral linewidth.

The strength of our inferred limit on the neutral fraction will depend on the choice of linewidth, as that determines the EW sensitivity. We discuss our choice of these values below in Appendix~\ref{app:inference_FWHM}.

The second term on the right-hand-side of Equation~\ref{appeqn:inference_linelike} can be expanded as:
\BE \label{appeqn:inference_linelike2}
p(W \,|\, \xHI, m, \mu, z_d) ={} \int_{-\infty}^{\infty} d\MUV \, p(W \,|\, \xHI, \MUV) p(\MUV \,|\, m, \mu, z_d) 
\EE
where the integral convolves the simulated $p(W \,|\, \xHI, \MUV)$ from \citetalias{Mason2018a} with the probability distribution of the absolute UV magnitude, $\MUV$, given our observed data:
\BE \label{appeqn:inference_likeWMuv}
p(\MUV \,|\, m, \mu, z_d) = \frac{1}{\sqrt{2\pi\sigma_M^2}}\exp{\left[-\frac{\left( \MUV - M_{\textsc{uv},\textrm{mod}}(\mu, z_d) \right)^2 }{2\sigma_M^2}\right]}
\EE
where $M_{\textsc{uv},\textrm{mod}}(\mu, z_d) = m - 5\log_{10}({D_L/10\,\textrm{pc}}) + 2.5\log_{10}(1+z_d) + 2.5\log_{10}{\mu}$ converts observed magnitudes to rest-frame UV magnitudes, assuming the UV spectral slope $\beta = -2$ to calculate the K-correction \citep[e.g.,][]{Blanton2007}. We assume the magnification distribution is log-normally distributed such that we can easily add the uncertainties in magnification and apparent magnitude: $\sigma_M^2 = \sigma_m^2 + (2.5\sigma_{\log{\mu}})^2$, where $\sigma_m$ is the error on the observed apparent magnitude and $\sigma_{\log{\mu}}$ is the uncertainty in the logarithmic magnifications. 

We note that the dependence on $\MUV$ in $p(W \,|\, \xHI, \MUV)$ is weak compared to the dependence on $\xHI$, and was parametrised with a smooth transition between two EW distributions for $\MUV > -20$ and $\MUV < -21$. Thus only the parameters which dominate changes in $\MUV$ are important. We note that the distance modulus term changes the magnitude in our redshift range of interest, $z=7.2-8.8$ ($\Delta \MUV \sim 0.5)$ less than magnification ($\Delta \MUV \sim 0.75$ for $\mu=2$), so for ease of computation we compute $p(W \,|\, \xHI, m, \mu, z_d)$ ahead of time for each galaxy, setting $z_d=8$, rather than having to redo this integral at every spectral pixel.

\subsection{Likelihood for a full spectrum}
\label{app:inference_like}

For a full spectrum $\{f\} = f(\lambda_i)$ the likelihood is just the product of the likelihoods at each wavelength pixel:
\BE \label{appeqn:inference_linelikefull}
\begin{split}
	&p(\{f\} \,|\, \xHI, m, \mu, z_d, \mathrm{FWHM}) =\\
	& \prod_i^N \int_0^\infty dW \, \frac{1}{\sqrt{2\pi}\sigma_i}e^{-\frac{1}{2} \left(\frac{f_i - f_\mathrm{mod,i}}{\sigma_i}\right)^2} p(W \,|\, \xHI, m, \mu, z_d)
\end{split}
\EE
where $d$ is the index of the emission line, and $z_d = \lambda_d/\lambda_\alpha - 1$, and $f_\mathrm{mod,i}$ is given by Equation~\ref{appeqn:inference_linemod}.

\subsection{Posteriors}
\label{app:inference_post}

Using Bayes' Theorem the posterior distribution for $\xHI$, FWHM and $z_d$ is
\BE \label{appeqn:inference_post}
\begin{split}
p(\xHI, z_d, \mathrm{FWHM} \,|\, \{f\}, m, \mu) \propto{}& p(\{f\} \,|\, \xHI, m, \mu, z_d, \mathrm{FWHM}) \\
& \times p(z_d)\,p(\xHI)\,p(\mathrm{FWHM})
\end{split}
\EE
We use a uniform prior on $\xHI$ between 0 and 1, we use the photometric redshift for the prior $p(z)$, and we use a log-normal prior on FWHM (discussed in Section~\ref{app:inference_FWHM}). As we are only interested in the posterior probability of $\xHI$ we can marginalise over $z_d$ and FWHM:
\BE \label{appeqn:inference_postmarg}
p(\xHI \,|\, \{f\}, m, \mu) = \int d\mathrm{FWHM} \; \int dz_d \; p(\xHI, z_d, \mathrm{FWHM} \,|\, \{f\}, m, \mu)
\EE
To account for the incomplete wavelength coverage, we make use of the fact if the object has \lya\ outside of the wavelength range (covering $[z_\textrm{min}$, $z_\textrm{max}]$) we would measure a non-detection in our data. Thus the integral over $z_d$ becomes:
\BE \label{appeqn:inference_postmargz}
\begin{split}
p(\xHI \,|\, \{f\}, m, \mu) \propto& \int_{z_\textrm{min}}^{z_\textrm{max}} dz_d \; p(\{f\}| \xHI, m, \mu, z_d) p(z_d) \\
			&+ \prod_i p(\{f\} = 0) \left( 1 - \int_{z_\textrm{min}}^{z_\textrm{max}} dz_d \; p(z_d) \right) 
\end{split}
\EE
We assume all galaxies observed are independent, so that the final posterior is the product of the normalised posteriors (Equation~\ref{appeqn:inference_postmarg}) for each object.

\subsection{The impact of linewidth on the inference}
\label{app:inference_FWHM}

\begin{figure}
\centering
\includegraphics[width=0.49\textwidth, clip=true]{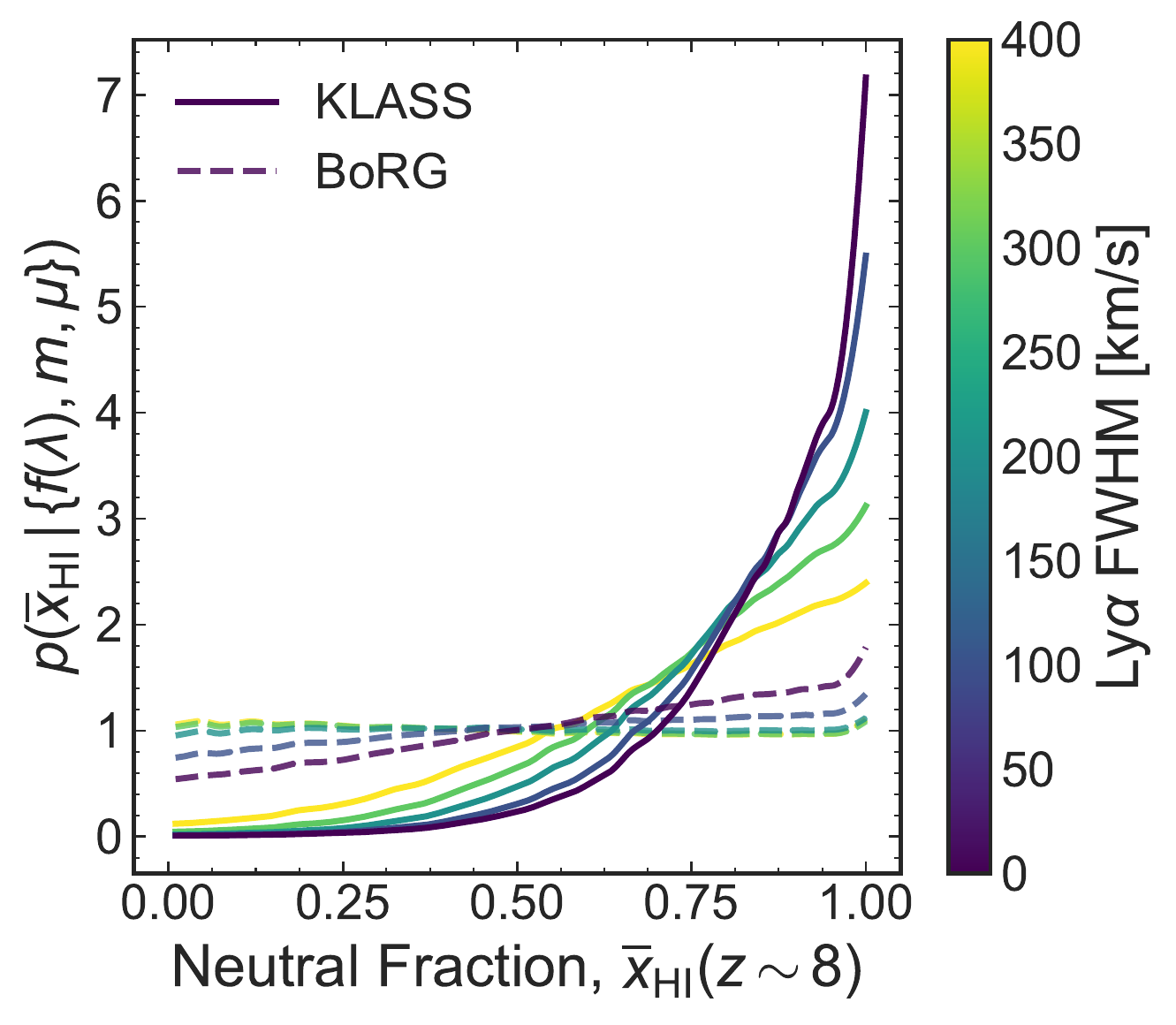}
\caption{Posterior probability distribution for the IGM neutral fraction derived using the KLASS (solid lines) and BORG (dashed lines) observations, as a function of the assumed FWHM (shown by the different colours) for the samples. As the assumed \lya\ FWHM increases, the inferred posterior flattens. This is because with increasing FWHM, our EW sensitivity decreases.}
\label{fig:app_post_compareFWHM}
\end{figure}

\begin{figure}
\centering
\includegraphics[width=0.45\textwidth, clip=true]{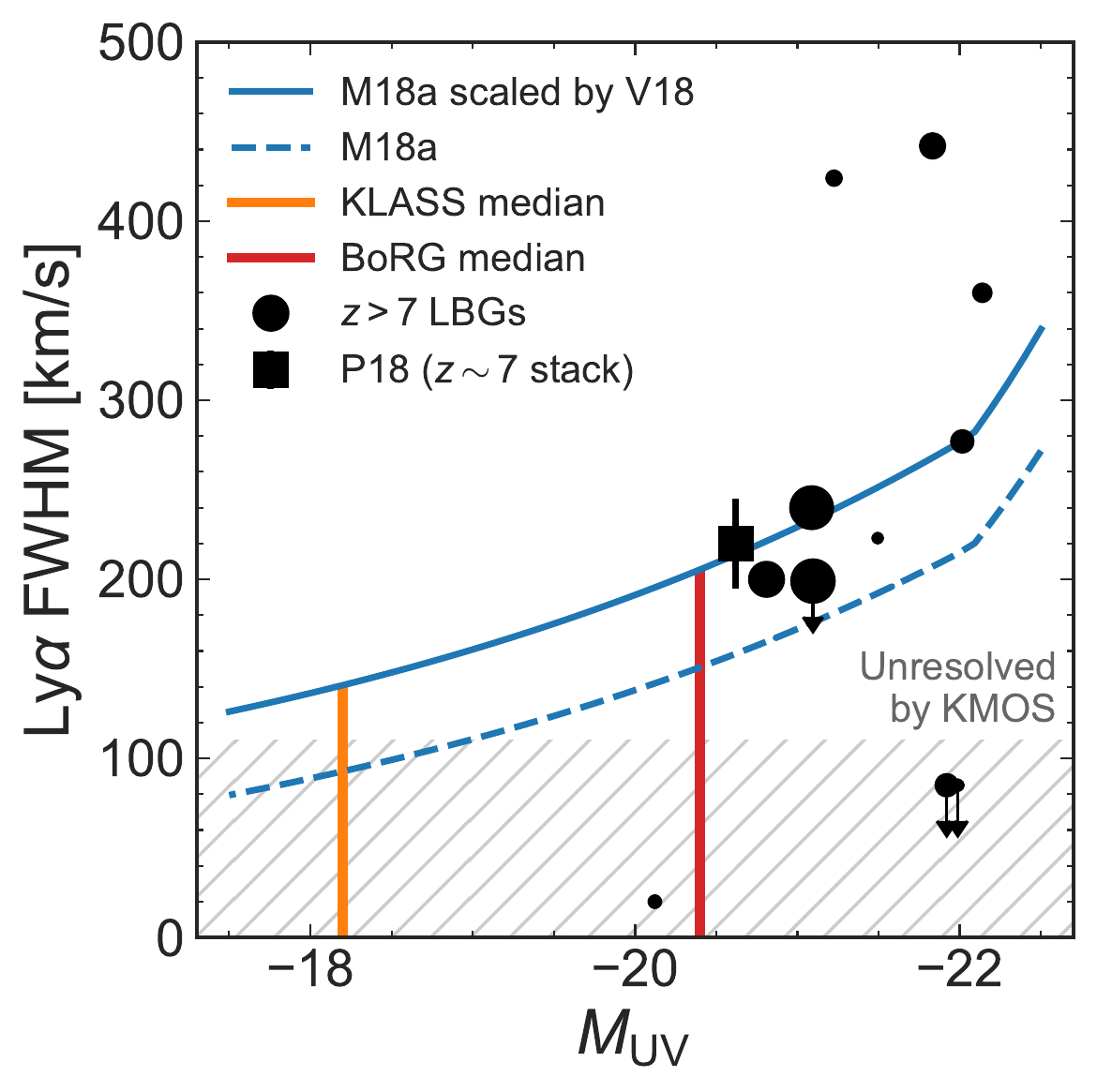}
\caption{\lya\ spectral FWHM as a function of UV magnitude. Black circles are measurements from LBGs with confirmed $z>7$ \lya\ \citep{Oesch2015,Roberts-Borsani2016,Stark2017,Zitrin2015a,Song2016,Finkelstein2013,Shibuya2012,Ono2012,Schenker2012,Vanzella2011,Laporte2017}. The black square shows median values derived from a $z\sim7$ sample by \citet{Pentericci2018}. The blue dashed line shows the \citetalias{Mason2018a}, model for velocity offset ($\DV$) as a function of UV magnitude and redshift (via halo mass). The blue solid line shows the \citetalias{Mason2018a} model scaled by the FWHM -- $\DV_{\mathrm{Ly}\alpha}$ relation presented by \citet{Verhamme2018}. The grey hatched region shows FWHM unresolved by KMOS ($\simlt 4$\,\AA\ or $\simlt 110$\,\kms). The orange (red) vertical line shows the median UV magnitude of the KLASS (BoRG) samples.}
\label{fig:app_post_FWHM_MUV}
\end{figure}

\begin{figure*}
\centering
\includegraphics[width=0.49\textwidth, clip=true]{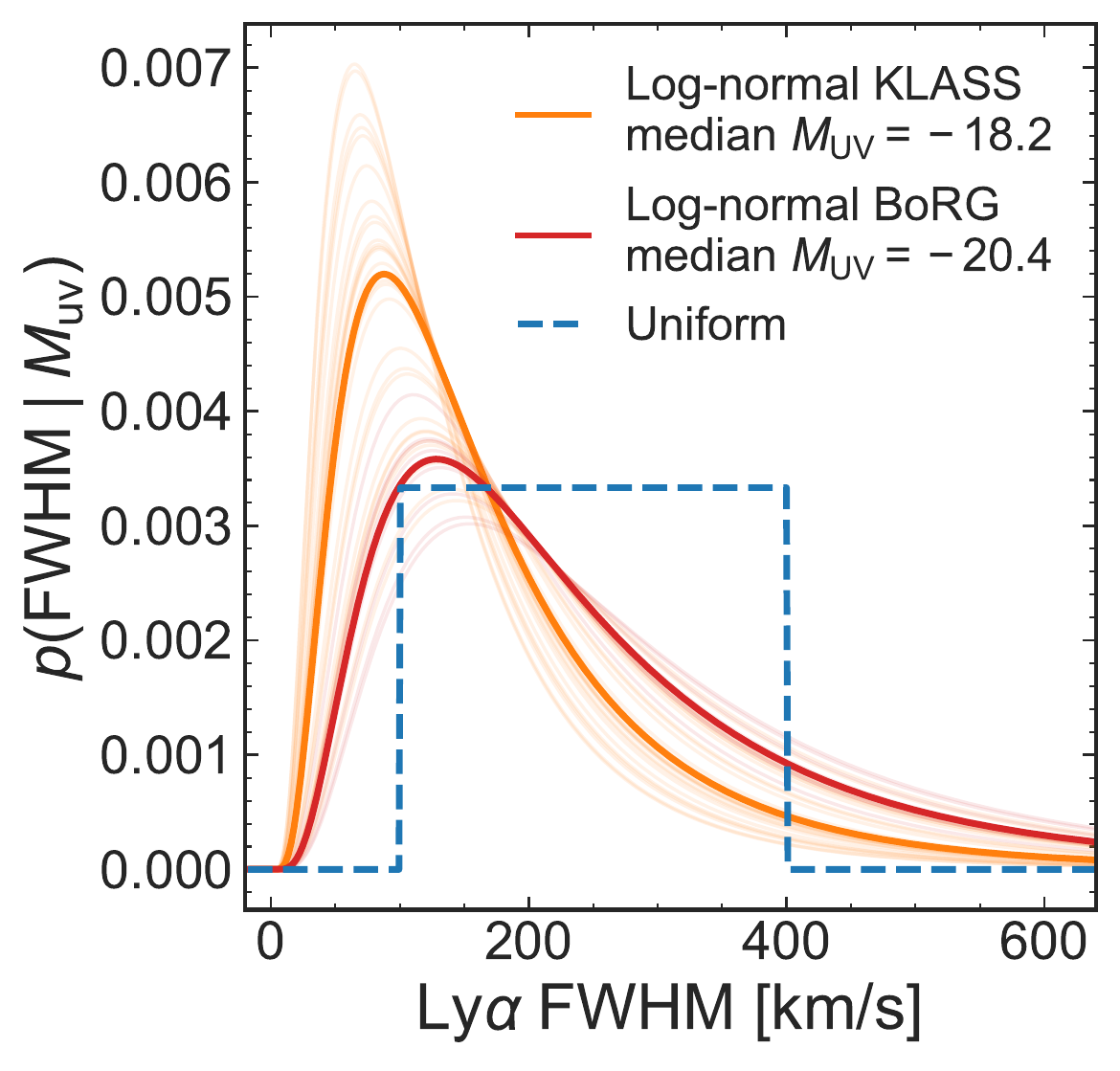}
\includegraphics[width=0.45\textwidth, clip=true]{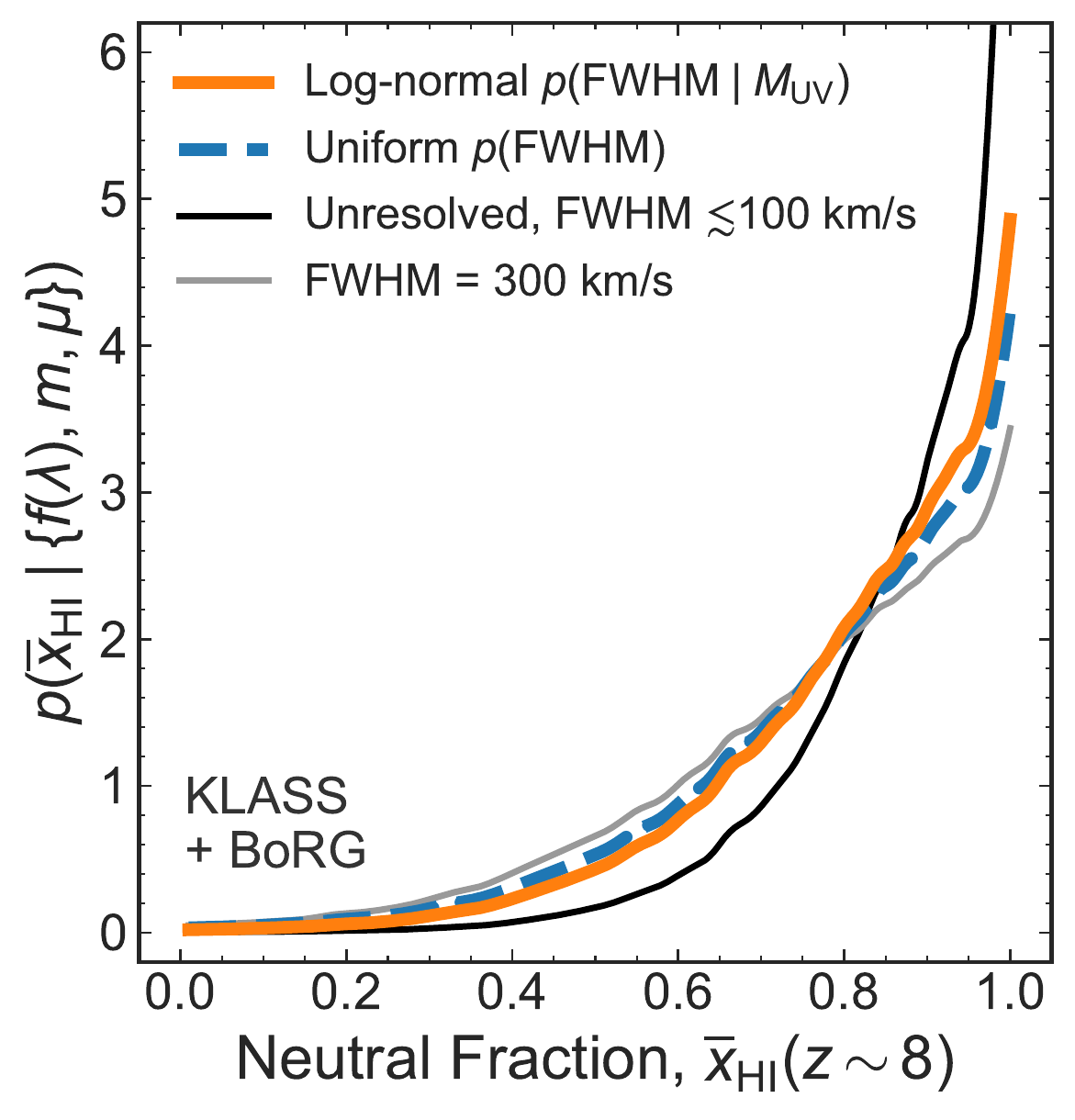}
\caption{\textbf{Left:} Example prior probability distributions for \lya\ FWHM used in our inferences. The orange curves shows the log-normal priors $p(\mathrm{FWHM} \,|\, \MUV)$ for the KLASS sample (median $\MUV = -18.2$ shown as thick orange line), while the red curves shows the log-normal priors for the BoRG sample (prior for median $\MUV = -20.4$ is the thick red line). The blue dashed line shows a uniform prior (independent of $\MUV$) between $100-400$\,\kms. \textbf{Right:} Posterior probability distribution for the IGM neutral fraction derived from the KLASS and BORG samples using several different FWHM priors. The thin black line shows the posterior obtained assuming all lines are unresolved (FWHM $\simlt 100$\,\kms) while the thin grey line is the posterior obtained assuming all emission lines are $300$\,\kms, demonstrating how the posterior flattens with increasing linewidth as the EW sensitivity decreases. For our UV faint targets we expect the lines to be relatively narrow, so the $300$\,\kms\ case is extreme. The orange solid line shows the posterior obtained using our log-normal FWHM prior $p(\mathrm{FWHM} \,|\, \MUV)$ calculated for each galaxy (described in Appendix~\ref{app:inference_FWHM}), and the blue dashed line shows the posterior obtained using the uniform prior which is independent of $\MUV$. The difference in the 68\% limits obtained using these two posteriors is negligible: $\Delta \xHI \sim 0.03$, Thus our results are robust for emission lines with widths $\sim100 - 400$\,\kms.}
\label{fig:app_post_comparepriors}
\end{figure*}

The observed linewidth of \lya\ emission will impact our inferences: broader lines will have lower significance and so reduce the strength of our inferences (see Figure~\ref{fig:app_post_compareFWHM}. As all of our observations are non-detections we must be careful to account for this effect, which we do by marginalising our posterior over FWHM using a empirically-motivated FWHM prior.

In Figure~\ref{fig:app_post_FWHM_MUV} we plot $\MUV$ versus \lya\ FWHM (deconvolved with instrumental resolution) for current detections of \lya\ in $z>7$ LBGs \citep{Oesch2015,Roberts-Borsani2016,Stark2017,Zitrin2015a,Song2016,Finkelstein2013,Shibuya2012,Ono2012,Schenker2012,Vanzella2011,Laporte2017}. The observed FWHM span $20 - 450$\,\kms. However linewidth measurements only exist for objects with $\MUV < -20$, significantly brighter than our KLASS sample, and there is a lot of scatter, motivating future observations to further explore the relationship.

Rather than fitting a relation to this limited current sample we construct FWHM priors using other simple empirical relations. To extrapolate to lower UV luminosities, motivated by observed correlations between \lya\ velocity offsets and the linewidths \citep{Verhamme2018}, we also plot the $\DV(\MUV)$ model derived empirically by \citet{Mason2018a} (their Equation 3). We then rescale this model using the relation between \lya\ FWHM and $\DV$ derived by \citep{Verhamme2018}: $\mathrm{FWHM} = (34+\DV)/0.9$. The \citet{Verhamme2018} relation is derived from a sample of $z\sim0-8$ \lya\ emitters, and suggests that the \lya\ photon scattering in ISM which produces this correlation between line offsets and linewidth is not entirely erased by the IGM.

We use two test FWHM priors: a uniform distribution between 100 and 400\,\kms, spanning the currently observed range; and a log-normal distribution with mean given by the \citet{Mason2018a} $\DV(\MUV)$ model scaled by the \citet{Verhamme2018} relation, with a 0.3 dex scatter. These priors are plotted in the left panel of Figure~\ref{fig:app_post_comparepriors}.

In the right panel of Figure~\ref{fig:app_post_comparepriors} we show the posteriors for the IGM neutral fraction obtained from the full KLASS and BoRG samples by marginalising over FWHM using these two priors, as well as the case of FWHM$=300$\,\kms\ and the case of unresolved lines. We see that expect for the case of unresolved lines the difference between the posteriors is negligible. Thus our results are robust for lines with FWHM$<400$\,\kms. We adopt the log-normal prior for our final inference as it is the most physically motivated.

If the lines are truly very narrow ($<100$\,\kms) our constraint would be stronger. However, without high-resolution observations of $z>5$ \lya\ in UV faint galaxies to provide evidence of narrow linewidths we have decided to be more conservative and use the log-normal prior in our final inferences.

\bsp	
\label{lastpage}
\end{document}